# Improved Non-parametric Penalized Maximum Likelihood Estimation for Arbitrarily Censored Survival Data

Justin D. Tubbs[1*], Lane G. Chen[1*], Thuan Quoc Thach [2], Pak C. Sham[1,2,3,#]

[1]Department of Psychiatry, The University of Hong Kong, Hong Kong, China

[2]Centre for PanorOmic Sciences, The University of Hong Kong, Hong Kong, China

[3]State Key Laboratory of Brain and Cognitive Sciences, The University of Hong Kong, Hong Kong, China

*These authors should be considered joint first authors.

[#]Corresponding author: pcsham@hku.hk 1/F HKJC Building for Interdisciplinary Research, 5 Sassoon Rd., Pokfulam, Hong Kong S.A.R., China

## Summary

Non-parametric maximum likelihood estimation encompasses a group of classic methods to estimate distribution-associated functions from potentially censored and truncated data, with extensive applications in survival analysis. These methods, including the Kaplan-Meier estimator and Turnbull's method, often result in overfitting, especially when the sample size is small. We propose an improvement to these methods by applying kernel smoothing to their raw estimates, based on a BIC-type loss function that balances the trade-off between optimizing model fit and controlling model complexity. In the context of a longitudinal study with repeated observations, we detail our proposed smoothing procedure and optimization algorithm. With extensive simulation studies over multiple realistic scenarios, we demonstrate that our smoothing-based procedure provides better overall accuracy in both survival function estimation and individual-level time-to-event prediction by reducing overfitting. Our smoothing procedure decreases the discrepancy between the estimated and true simulated survival function using interval-censored data by up to 49% compared to the raw un-smoothed estimate, with similar improvements of up to 41% and 23% in within-sample and out-of-sample prediction, respectively. Finally, we apply our method to real data on censored breast cancer diagnosis, which similarly shows improvement when compared to empirical survival estimates from uncensored data. We provide an R package, SISE, for implementing our penalized likelihood method.

**Introduction**

Estimating time-to-event distributions and their associated survival functions is important in demography, epidemiology, clinical trials, and other fields. The classic Kaplan & Meier (1958) product limit estimator (KM) is widely used to estimate survival functions from potentially right-censored observations with

$$\hat{S}(t) = \prod_{j:t_j \leq t} \left(1 - \frac{d_j}{n_j}\right) \quad (1.1)$$

where $0 \leqslant t_1 < t_2 < \cdots < t$ represents distinct times, $d_j$ represents the number of events observed at time $t_j$, and $n_j$ represents the number of individuals known to have survived up to time $t_j$. The estimator is a monotonic step function, decreasing by the discrete hazard rate ($\frac{d_j}{n_j} \geq 0$) that is the fraction of subjects at risk who experienced the event at $t_j$. The estimated survival function can be transformed to give an estimate of the probability mass function $f(t)$ by

$$\hat{f}(t) = (1 - \hat{S}(t)) - (1 - \hat{S}(t-1)) \quad (1.2)$$

The KM estimator has been shown to be a non-parametric maximum likelihood estimator (NPMLE) (Peterson, 1977) and is widely used in biomedical research.

Turnbull (1974) improved Peto's method (1973) and proposed a generalization of the KM estimator, where each individual's time-to-event is not measured exactly, but only to be contained in a certain time interval. For example, interval censoring occurs when subjects are assessed periodically at particular time points, so that each subject's time-to-event can be inferred only to be within the interval between the last measurement timepoint before event occurrence ($L_i$) and the first timepoint after ($R_i$). The algorithm is iterative, where the contributions of an observation to survival are updated according to the overall survival function estimated at the previous iteration. TB is the standard nonparametric procedure for obtaining ML estimates of survival functions from arbitrarily censored and truncated data.

However, NPMLE, including KM, TB and the EM algorithm for incomplete data (Dempster et al., 1977), commonly faces the problem of overfitting when applied to small samples, resulting in poor estimation (Bishop, 2016). Under certain regularity conditions, NPMLE for density function estimation can be reduced to fitting a finite mixture model with at most N components (Feng & Dicker, 2018). Thus, overfitting can easily occur when the model complexity is unbound and the number of components is unrestricted. This issue is particularly serious when the true density of a finite sample itself is continuously differentiable but available sample size is insufficient.

Two major approaches have been considered to address this issue. The first approach is moving to a semi-parametric paradigm by restricting the family (density function form) of each component. A classic example of this approach is the Finite Mixture Gaussian Models, which approximates a non-parametric model as the number of components becomes large. Kuo & Peng(1995) applied EM and Monte Carlo EM to fit Mixture Models for survival data. Other attempts in survival analysis include spline-based approaches (Yavuz & Lambert, 2011; Zhang et al., 2010). However, automatic hyperparameter tuning, computational infeasibility and non-convergence issues remain common challenges to these semi-parametric procedures. The second approach is smoothing. Smoothing NPMLE to reduce overfitting is intuitively reasonable because a poor NPMLE estimate of a density curve tends to have undesirably high variability. Such density estimates may be non-smooth with non-differentiable points, be greatly fluctuating with many ups and downs, be too "spiky" with steep peaks and troughs, and possibly even contain many regions of non-continuity. Kernel density estimation (KDE), is a fundamental non-parametric smoothing approach that treats each data point as a distribution of kernel density with the same form (Parzen, 1962; Rosenblatt, 1956; Whittle, 1958). It is commonly used to correct histograms constructed from small samples, which tend to overfit the data. Smoothing by KDE is controlled by both the type of kernel and the window size (bandwidth). KM and TB could be regarded as an NPMLE analogue of the histogram for censored data, so it is very natural to consider applying smoothing as a simple solution to excessive variability of the density estimate. A recent example in survival analysis used saddle point approximation to smooth survival and hazard function (Dissanayake & Trindade, 2020), but has no convenient implementation in common statistical software.

A crucial issue for either approach is how to select the most parsimonious model that fits the data well. For Gaussian Mixtures, the key is choosing the optimal number of components (Melnykov & Maitra, 2010). In terms of smoothing, it is to determine the optimal degree of smoothing to be applied, i.e. choosing the smoothing parameter (Chu & Marron, 1991). Information Criteria (IC) are a simple but fundamental class of methods for model selection. Based on Information Theory or Bayesian model selection principles, an information criterion usually penalizes the log-likelihood by a measure of model complexity, capacity or dimensionality. For example, Akaike information criterion (AIC) (Akaike, 1974) provides an unbiased estimate of the expected KL divergence of the selected model from the true model under certain assumptions, which minimizes the information loss and maximizes the "entropy":

$$AIC = -2\ln L + 2k \qquad (1.3)$$

where L is the maximized likelihood of the selected model, k is the corresponding number of parameters, and N is the sample size. Another commonly used IC is the Bayesian Information Criterion (BIC) (Schwarz, 1978) that approximately maximizes the posterior probability of the selected model with assignment of equal prior probability for each candidate

$$BIC = -2\ln L + k \times \ln N \qquad (1.4)$$

Although IC were initially proposed for parametric model selection, the idea is generalizable to non-parametric cases by identifying an analog of the number of parameters. For example, Hurvich, Simonoff and Tsai (1998) proposed the $AIC_c$, which treats the trace of the smoother matrix as the effective number of parameters for any linear smoother, which can reduce undesirable variability and tendency to under-smooth. For smoothing, IC are generally preferred over existing alternative methods such as Cross Validation and Plug-in Estimator (Schucany, 2004), although there is considerable debate as to what procedure gives superior performance under various situations (Jones et al., 1996; Loader, 1999).

In this paper, we propose a straightforward kernel smoothing procedure for KM and TB estimates to obtain more accurate estimates of survival and probability density functions. The procedure uses a modified BIC to penalize the log-likelihood function by the degree of variability in the probability density function. In contrast to previously proposed methods, our approach requires no tuning parameter selection and is easily implemented in an R package. We show that the proposed procedure results in substantial improvements in estimation accuracy in a wide range of realistic scenarios, by simulation studies and applications to real data.

**Method**

**Data Specification and Censoring Representation**

We follow the notation employed by Turnbull (1976) and illustrate our proposed algorithm in the familiar context of disease onset estimation using longitudinal follow-up data. The random variable of interest is the time-to-onset (event) $\{X_1 = x_1, ..., X_N = x_N\}$ ($X_i \geq 0, 1 \leq i \leq N$) measured in N independent subjects. In a disease onset model, c represents the earliest time at which an individual is at risk for a disease, and C represents the latest time of potential risk. These can be adjusted based on evidence or hypotheses about the underlying characteristics of the process under study.

Let $\{X_1, \ldots, X_N\}$ be independently distributed (*iid*) from $f_X(x)$ with support $(c, C) \subset (0, +\infty)$. Here we treat time-to-event as continuous for convention. We aim to estimate the density function $f_X(x)$ and thus its associated survival function $S(t)$.

We also assume that the measured event is "irreversible", such as death or chronic disease onset. For each individual, the observed data contain $m_i$ pairs of observations $(Z_{ij}, T_{ij}), 1 \leq j \leq m_j$, where $Z_{ij}$ indicates the disease status, 0 or 1, of individual i at observation time $T_{ij}$.

For TB estimation, this original $N \times m$ data matrix can be summarized into N censored data intervals $A_i = (L_i, R_i), 1 \leq i \leq N$ with $L_i < X_i < R_i$, where "ties" of individuals with the same interval bounds are possible. With this two-number summary, we are able to express left censoring, interval censoring, and right censoring data types. Table 1 illustrates a hypothetical dataset concerning heart attack history of four individuals with three observations each, showing the originally observed data and its summarized forms. Considering individual 1, we note that the subject had never suffered a heart attack by age 38, but reported a confirmed history upon follow-up at age 60 (and therefore 70). Thus an investigator is able to ascertain that the individual's first heart attack occurred between age 38 and 60, i.e. $38 < x_1 < 60$. Similarly, $41 < x_2 < 48, x_3 > 62, x_4 < 36$. *Summarized data* are a group of "intervals" as many as sample size, likewise $\{(38,60), (41,48), (62, +\infty), (0,36)\}$.

We can transform any left-censored data $(-\infty, R_i)$ into interval censored data $(0, R_i)$ or a narrower $(c, R_i)$ if we know that the left boundary c is greater than zero. When $R_i$ is finite, $X_i$ is interval censored; when $R_i = +\infty$, $X_i$ is right censored. With a known finite C, we can similarly transform any right-censored data $(L_i, +\infty,)$ into interval censored data $(L_i, C)$.

**Representation of Truncation**

In practice, every time-to-event data point is potentially truncated by a set $B_i = (B_i^L, B_i^R)$ representing the time range of observable events. Any event outside the bounds of B could not be measured or recorded. Truncation often occurs when the event has been subject to some screening procedure so that all observations outside truncation sets have been removed. Right truncation is very common because it can occur due to exit from study, death of research subject, end of follow-up, etc. If at the last observation, the individual still has not experienced the disease, the onset may either occur later, or never occur before death, being non-observable. For this reason, it is only possible to estimate a survival curve up to the time when the last observation occurred, and anything beyond this time is pure extrapolation.

Although truncation sets are individual-specific, for simplicity we can assume non-differential truncation $B_1^L = B_2^L = \cdots = B^L$ and $B_1^R = B_2^R = \cdots = B_N^R = B^R$. This means that all individuals share a common truncation set. We call this common set $B = (B^L, B^R)$ the "event occurrence time frame", with $B^L$ and $B^R$ being the left and right boundaries of the time range of any possibly observed event. In practice, like the support of the full density $(c, C)$, $(B^L, B^R)$ is often unknown, but we can approximate them with the time of the earliest and latest observation (the "observed time frame") in the study.

$$\widehat{B^L} = \min_{ij} T_{ij} \geq 0, \widehat{B^R} = \max_{ij} T_{ij} \tag{2.1}$$

With an appropriate etiologically-informed study design, such as a birth cohort for congenital diseases or a cohort beginning in late adulthood for geriatric diseases, the study period would be good approximation to the common risk-observable period.

**Relation to Prevalence and Interpretation**

Under our notations, the lifetime prevalence of a disease in the population is considered to be

$$p = \int_0^{B^R} dF_X(x) \tag{2.2}$$

Which is usually approximated by the empirical prevalence in the cohort

$$\hat{p} \approx \int_0^{\widehat{B^R}} d\hat{F}_X(x) \tag{2.3}$$

It is associated with survival probability at time $B^R$ as

$$S(B^R) = \int_{B^R}^C dF_X(x) = 1 - p \tag{2.4}$$

Thus, the truncated portion of the full density $f_X(x)$ represents those individuals who would never develop the disease within their lifetime. The TB method is able to handle the complicated issue of truncation adequately by providing an estimator for the truncated distribution $F_X(x; B) = P(X \leq x | X \in B)$ instead of the full distribution. In reality, $(c, C)$ and $(B_L, B_R)$ are not guaranteed to greatly overlap with each other. For rare diseases, $C$ would be very large or may even be $+\infty$, and thus many possible disease onset events are truncated from real study. But if $C$ is close to $B^R$, say 95 years, then lifetime prevalence for such a disease, like certain type of common cancer, is close to one. Anyone who lives long enough would be expected to develop the disease at some time. In this case, right truncation is not a great issue.

**Estimation Procedure and Constrained Optimization**

As shown in Supplementary Figure S1, the raw TB estimate of the probability density function is over-fitted and has excessive variability compared to the true underlying density, with peaks and troughs becoming discontinuous "spikes". Applying kernel smoothing to raw TB is a simple and natural solution. However, from Supplementary Figure S2 we can see that under-smoothing (with small window size) can introduce continuity and some smoothness, but may not overcome the excessive variability completely, whereas over-smoothing could lead to undesired underfitting instead. Naturally, kernel smoothing is more effective with small sample sizes and therefore more preferred because the observations provides insufficient information for accurate NPMLE result. As the sample size increases, the NPMLE itself will have better performance by avoiding overfitting and thus requiring less smoothing.

The major purpose of smoothing is to reduce model complexity, reflected by two aspects: dimensionality (e.g. the number of independent variables) and variability (the order of fitted polynomial). Classic BIC seeks a balance between parsimonious model complexity and high log-likelihood, but its analog in non-parametric contexts is usually unclear. Inspired by $AIC_c$ (Hurvich et al.,1998) and the classic BIC, we propose a loss function of similar form for non-parametric estimation, the "smoothing BIC":

$$BIC_s = -2\ln L + k_T \times \ln N_s \qquad (2.5)$$

We choose to penalize on the number of "turning points" ($k_T$) of $\hat{f}_n$, the resulted density function estimate (either from NPMLE or after smoothing). The penalty is weighted by the natural logarithm of a sample-based information quantity $N_s$, so that the severity of the penalty term will decay rapidly with increasing sample size. We define $k_T$ as the number of "turning points", i.e., how frequently the gradient of the density function estimate changes sign. This may be thought of as a measure of the variability or complexity of a given density function estimate. An inverted U or V shape would contribute one to $k_T$. It is almost identical to *the number of strict local extrema*, except that a "semi-strict" local extrema (like a slope step shape ⌐ ) would also contribute one to $k_T$ instead of zero. Mathematically, under the condition that $\hat{f}_n(x)$ is continuously differentiable, $k_T$ equals the number of jump discontinuity points of function $\text{sign}\{\hat{f}'_n(x)\}$, which can be expressed as

$$k_T := \text{card}(\{x \in \mathbb{R} | \text{sign}\left(\hat{f}'_n(x^+)\right) \neq \text{sign}\left(\hat{f}'_n(x^-)\right)\}) \qquad (2.6)$$

Intuitively, a smooth curve with n local extrema can be approximately fitted by a polynomial of order n+1 (or higher), which can by characterized by n+2 coefficients. Say $P_{n+1}(x) = a_0(x - a_1) \ldots (x - a_n)(x - a_{n+1}) = P_n(x)(x - a_{n+1})$, then equation $f'(x) \approx$

$P'_{n+1}(x) = P_n(x) + P'_n(x)(x - a_{n+1}) = 0$ could have n distinct roots, representing n local extremum points as we desire. Therefore, the number of turning points could be considered a reasonable analog of the number of parameters under the density function estimation scenario.

In programming, the value of $k_T$ can be approximated by calculating finite differences between neighboring small bins. We avoid counting trivial hyper-local changes in sign by only considering those where the absolute difference between neighboring bins is less than 1% of the average neighboring bin difference across all bins.

The $\ln N_s$ component of our $BIC_s$ may take multiple forms. In our later simulations, we consider three different alternatives of $\ln N_s$:

$$\text{(the sample size) } \ln N := \ln(N)$$

$$\text{(the number of observations) } \ln N_m := \ln \sum_i m_i \ (= \ln(N \times m))$$

$$\text{(equivalent sample size) } \ln N_e := \ln(N_e)$$

We may notice that for interval-censored data, a wide interval censored observation is usually less informative than a narrow interval. With a "width" of zero, uncensored observation would be the most informative. Thus, to find an estimate of the sample size equivalent in the sense of non-censoring, we define a new quantity $N_e$ based on the informativeness of the observed sample

$$N_e := \sum_i^N n_i$$

$$n_i := 1 - \left(S(L_i) - S(R_i)\right) \approx 1 - \left(\hat{S}(L_i) - \hat{S}(R_i)\right) \tag{2.7}$$

where $\hat{S}$ is the estimated survival function.

$N_e$ attempts to quantify the density of observations within the preferred overlap of the "event occurrence time frame" $(B^L, B^R)$ and the "observed time frame" (actual study period) $[\min_{ij} T_{ij}, \max_{ij} T_{ij}]$ by accounting for both the positioning of the time-to-event intervals and their precision (width). We present a visualization of some properties of $N_e$ in the supporting information.

Applying the Nadaraya-Watson (NW) normal kernel regression smoothing procedure (Nadaraya, 1964; Watson, 1964) to the distribution density estimated using the *interval* package in R (Fay & Shaw, 2010), we aim to find an optimal window size d which minimizes the $BIC_s$, while constraining d to be non-negative and less than $B^R \times (t_2 - t_1)$ to avoid unreasonably large window sizes. An advantage about NW kernel regression is that weighted NW estimators are always non-negative if non-negative kernels are used. Additionally, we found from small-scale simulation optimization that NW estimation achieved slightly better overall performance and used less computing resources, however it is possible that other smoothing procedures may also work well.

Following the definitions in Turnbull(1976) and Gentleman & Geyer(1994) the log-likelihood for interval-censored data is

$$\ln L = \sum_i^N \ln f(L_i, R_i) \qquad (2.8)$$

$$f(L_i, R_i) = \begin{cases} \dfrac{f_X^{(d)}(t_i)}{P_F(B_i)}; & \text{if } L_i = t_i = R_i \\ \dfrac{F(R_i^+) - F(L_i^-)}{P_F(B_i)} = \dfrac{\int_{L_i}^{R_i} f_X^{(d)}(t)dt}{P_F(B_i)}; & L_i < R_i \end{cases} \qquad (2.9)$$

where $f_X^{(d)}(t)$ is the estimated density function obtained from smoothing the raw TB estimate, $\hat{f}_X^{(TB)}(t)$ with normal kernel of window size d.

With the assumption all $B_i = B$,

$$P_F(B_i) = p. \qquad (2.10)$$

Furthermore, if $B = (0, +\infty) \supset (c, C)$, then

$$P_F(B_i) = 1. \qquad (2.11)$$

In general, as shown in Figure 1, as d increases, the $-2\ln L(d)$ increases while the penalty component $k_T(d)\ln N_s$ decreases. Thus, our objective is to find a global optimum d that can achieve the best trade-off in terms of BIC$_s$.

To solve this optimization problem, we employ global-local optimization algorithms implemented in *Nlopt* (Johnson, n.d.; Ypma et al., 2020), an open-source library for nonlinear optimization. Specifically, we choose ISRES (Improved Stochastic Ranking Evolution Strategy) (Runarsson & Yao, 2005) for global optimization with a starting point of zero, and Nelder-Mead (Nelder & Mead, 1965) simplex algorithm for local optimization. We reserved the first local optimum found as our solution to avoid unnecessary computational burden. Although theoretically there could a situation where another window size yields similarly minimal BIC$_s$, we expect the solution to be very close to globally optimal window size, since our inspection of BIC$_s$ properties at Figure 1 revealed a generally convex form without strong sensitivity. We provide an R package, called SISE (https://github.com/tubbsjd/SISE) which implements our algorithm in a single convenient function. Here we summarize our overall estimation process beginning from *summarized data*.

**Algorithm Summary:**

1. Obtain the traditional non-parametric (TB or KM) estimate of survival (and thus, time-to-event) distribution.
    a. If employing the $N_e$-based penalty, calculate the $N_e$ assuming the distribution estimated in step 1.
2. Apply the NW normal kernel regression smoother with window size d to the time-to-event distribution estimated in step 1.
3. Given the smoothed distribution calculated from step 2, calculate the empirical log-likelihood, the number of "turning points" on the curve, the sample size-informed quantity $N_s$ and thus $BIC_s$ ($= -2\ln L + k_T \ln N_s$) of the data.
4. Repeat steps 2 and 3, varying values of window size d, updating $BIC_s$, until convergence to a smoothed estimate which minimizes the $BIC_s$.

**Data Simulation and Sampling Scheme**

Our primary simulation procedure (S1) tests the performance of our proposed smoothing procedure across multiple scenarios, reflecting varied underlying disease models and sampling study designs. Secondarily, S1 compares the performance of different options for the $\ln N_s$ component of our $BIC_s$ penalty and the application of our proposed smoothing procedure to traditional KM estimation results. For S1, we performed 100 replicates for each of 108 parameter combinations, varying the sample size (N), mean onset age (u), lifetime prevalence (p), and number of repeated observations per person (m). In simulation 2 (S2), we tested the robustness of our method to an underlying disease process which is bimodal by performing 500 replicates under a single parameter configuration. Finally, in simulation 3 (S3), we demonstrate the performance of bootstrapping to estimate smooth confidence intervals in 100 replications of a single parameter configuration. A summary of these simulations and their parameters is given in Table 2.

In general, across all simulations, we generate realizations of $\{X_1, ..., X_N\}$, with precision to two decimal places, from a true distribution density $f_X(x)$. The values stand for the time-to-onset (event occurrence) of N individuals in a study. Each $X_i$ can be conceptualized as one's age at disease onset if ages of all individuals have been aligned up.

For simplicity, we assume X to follow the *mixture distribution of log-normal and a $(1-p)$ point mass at a very large number K* ($K \gg B^R$) in simulation such that

$$\ln X_i | H_i \sim H_i \times N(\mu, \sigma^2) + (1-H_i) \times \delta(K)$$
$$H_i \sim \text{Bernoulli}(p) \tag{2.12}$$

The point mass, with a probability of one minus the lifetime prevalence, represents the proportion of individuals who never developed the disease for lifetime. The value of K can be arbitrary as long as it is large enough compared to $B^R$.

Therefore, we have

$$X_i | H_i = 1 \sim \text{logN}(u, s^2) \tag{2.13}$$

with mean u and standard deviation s such that

$$\mu = \ln \frac{u^2}{(u^2 + s^2)^{\frac{1}{2}}}, \sigma^2 = \ln\left(1 + \frac{s^2}{u^2}\right) \tag{2.14}$$

Obviously, the support $(c, C) = (0, +\infty)$. For all simulations, we fix $s = 10$. For simplicity, we fix the common truncation set B as $(0, B^R)$, with $B^R = 100.00$ to limit the maximal age at possible disease onset to 100 years old, so K can be like 1000.

Subsequently, we generate originally observed data $\{(Z_{ij}, T_{ij}), 1 \leq i \leq N, 1 \leq j \leq m\}$, where m is the number of repeated observations (at different time points) for each individual.

We base our sampling scheme on truncated normal distributions. Specifically, we first generate a set of baseline observation times, sampling from a *normal distribution* and truncating it by $(0, B^R)$

$$T_{i1} \sim N(\mu_1, \sigma_1^2) \cap 0 < T_{i1} < B^R \tag{2.15}$$

Second, we generate a set of time gaps to the next observation by sampling from a *normal distribution* and then truncating it by $(0, +\infty)$ to ensure positive follow up times

$$\Delta T \sim N\left(\frac{L}{m-1}, \sigma_\Delta^2\right) \cap \Delta T > 0 \tag{2.16}$$

where L is the average length of the whole follow-up period and $\frac{L}{m-1}$ is the average time to next follow-up.

Then, we generate the next observation time by adding the last observation time with the time gap and performing truncation with $(0, B^R)$

$$T_{ij} = T_{i(j-1)} + \Delta T, 2 \leq j \leq m \cap 0 < T_{ij} < B^R \tag{2.17}$$

Along with our simulation settings, the expected "observed time frame" is approximately $[u - e^{\mu(u,s) - 3\sigma(u,s)}, L + 3\sigma_\Delta]$. We fix $L = 20$, representing an average total study period of 20 years. Further, we fix $\sigma_1 = 10, \sigma_\Delta = 0.2$. The mean age at first observation $\mu_1$ is fixed to 40 for all simulations, therefore $\mu_2 = 40 + 20, \mu_3 = 40 + 2 \times 20$. Under such circumstances, a sampling procedure would be considered appropriate when the true mean onset age u is around 50, but when u is truly 30 or 70, the main study period would be too late or too early to capture the majority of the onset occurrences.

Based on the realized values of $T_{ij}$ and $X_i$, we infer the disease status at the $j^{th}$ observation for individual i at time $T_{ij}$

$$Z_{ij} = 1_{\{T_{ij} \geq X_i\}} \quad (2.18)$$

Raw data of $N \times m$ pairs $\{(Z_{ij}, T_{ij}), 1 \leq i \leq N, 1 \leq j \leq m\}$ are then transformed to two-term *summarized data* $\{(L_i, R_i), 1 \leq i \leq N\}$, defining

$$L_i = \max_j\{T_{ij} * (1 - Z_{ij}) + c * Z_{ij}\} \quad (2.19)$$

Assuming $c = 0$, then $L_i = \max_j\{T_{ij} * (1 - Z_{ij})\}$

Similarly,

$$R_i = \min_j\{T_{ij} * Z_{ij} + C * (1 - Z_{ij})\} \quad (2.20)$$

In other words, $L_i$ is the time-point at which the $i^{th}$ individual was last observed to be disease-free, or equals to c if the individual had already developed the disease at the first observation (i.e. all $Z_{ij}$ are one). Similarly, $R_i$ is the time-point at which the $i^{th}$ individual was first observed to have the disease, or equals C if the individual has not developed the disease, even at the last observation (i.e. all $Z_{ij}$ are zero).

**Evaluation and Comparison**

For S1 and S2, we evaluated our smoothing method performance using three aspects. Similar to the Mean Integrated Squared Error (MISE) commonly used in density estimation, we evaluate the Average Root Integrated Absolute Error (ARISE) of how much the estimate deviates from the true survival function

$$\widehat{RISE} = \frac{1}{p} \times \sqrt{\frac{\sum_{j=1}^{l^{(r)}} \left(\hat{S}^{(r)}\left(\tau_j^{(r)}\right) - S\left(\tau_j^{(r)}\right)\right)^2}{l^{(r)}}} \quad (2.22)$$

$$\widehat{ARISE} = \frac{\sum_{r=1}^{M} \widehat{RISE}}{M} \quad (2.23)$$

where $S(\cdot)$ is the true survival function which is known in simulation and $S^{(r)}(\cdot)$ is the estimated survival function based on the $r^{th}$ set of $\{(Z_{ij}, T_{ij}), 1 \leq i \leq N, 1 \leq j \leq m\}$ data simulated from the same model $\{f_X(x), g(T_1, \ldots, T_m), p\}$. $\tau_j^{(r)}, 1 \leq j \leq l^{(r)}$ are sequential distinct time points from $\min\{T_{ij}^{(r)}\}$ to $\max\{T_{ij}^{(r)}\}$ with step size $\delta t = \tau_j - \tau_{j-1}$ as the minimal decimal unit (1, 0.1, 0.01 etc.) of the data T. Thus in evaluation we don't consider the extrapolation from $(0, \min\{T_{ij}^{(r)}\})$ but only measure accuracy within the observed data range. Here $\delta t = 0.01$. The total number of time points $l^{(r)} = \frac{\max\{T_{ij}^{(r)}\} - \min\{T_{ij}^{(r)}\}}{\delta t}$. We adjust the RISE to account for the influence of lifetime prevalence p by dividing RISE by p. This is done to standardize the maximum deviation so that it is still reasonable to make comparison of models across configurations where p is potentially different. The geometric interpretation of $\widehat{\text{RISE}}$ is the enclosed area between estimated survival curve $\hat{S}(t)$ and the true curve $S(t)$ divided by the total rectangular area $(\max T_{ij} - \min T_{ij}) \times p$.

Additionally, we measure the Average Root Mean Squared Error (ARMSE) for "recovering" exact event occurrence time. In $r^{th}$ simluation, $\mathbf{X}^{(r)} = \{x_1, \ldots, x_N\}(x_i \geq 0, 1 \leq i \leq N)$ is potentially censored, so we impute $\mathbf{X}^{(r)}$ by $\hat{\mathbf{X}} = \{\hat{x}_1, \ldots, \hat{x}_N\}$, whose element is the expectation in the interval based on the estimated density, approximated by a weighted sum.

$$\hat{x}_i = E(X_i | X_i \in (L_i, R_i)) = \int_{L_i}^{R_i} x \, \widehat{f_X}(x) dx \quad (2.24)$$

$$\text{ARMSE} \approx \frac{\sum_{r=1}^{M} \text{RMSE}}{M} \quad (2.25)$$

$$\text{RMSE} = \sqrt{\frac{\sum_{k=1}^{N} (\widehat{X_k} - x_k)^2}{N}} \quad (2.26)$$

In S1 and S2, we measure the performance of prediction using ARMSE both within the sample used for survival estimation (ARMSEw) and in an external sample (ARMSEo), applying our imputation procedure only to interval-censored individuals. To test out-of-sample prediction, we simulated an independent dataset with the same configuration parameters as the sample used to estimate the survival curve. In practice, it is possible that the un-smoothed TB density estimator contains regions of zero probability mass in an out-of-sample interval. Thus, in order to make imputation in these scenarios possible, we inflated any regions of zero probability to a very small value close to zero during the imputation process.

For S3, we obtained raw and smoothed TB estimates for 200 bootstrapped samples at each independent simulation replicate, allowing us to approximate 95% confidence intervals for each age band by the 0.025 and 0.975 quantiles of the 200 bootstrapped estimators. For each replicate, we determined the proportion of age bands where the confidence interval contained the true survival estimate.

For S1 and S2, in addition to applying our proposed smoothing routine to the survival function as estimated by the TB algorithm using censored data, we also tested its performance on the results from traditional KM estimation as implemented in the survival package in R (Therneau, 2021; Therneau & Grambsch, 2000). In this scenario, the data provided to the estimation procedure was the vector of simulated exact individual onset ages if this age is before the individual's last measurement. If an individual's onset age falls after their final observation or they were assigned to the point mass of individuals who will never develop the disease, they were considered right censored at their age of last observation. For KM smoothing in S1, we only applied smoothing using the lnNe-penalized $BIC_s$.

**Real Data Application**

To demonstrate the performance and applicability of our smoothing method in real situations, we utilized data on breast cancer diagnosis (BC) in the publicly available Midlife in the United States (MIDUS) dataset. For a full description of the cohort, see Radler (2014). Briefly, a group of around 7,000 American adults were followed for up to three time points where they answered questions via telephone related to their health and lifestyle, subsequent studies have added to this original cohort and collected additional measures. For each individual in the original cohort, we obtained their reported current age, year of birth (YOB), breast cancer diagnosis status, and year of breast cancer diagnosis (only assessed at two time points). Thus, this dataset provides both traditional binary interval-censored BC onset data and a more exact self-reported time of onset.

We first filtered the data to retain only females and those who had complete data for at least one time point. For those who reported a BC diagnosis, we derived their reported age at onset (BCage) from their YOB and year of diagnosis information. If they had reported BCage at more than one time point, the average of all reports was taken. Additionally, we removed any individuals who had an absolute difference of greater than 5 years between two reports of BCage, or those who reported a BC diagnosis, but did not report a BCage. Finally, we excluded individuals who had inconsistent binary reports of previous BC diagnosis (i.e. reporting BC at one time point, and reporting never having BC at subsequent time point), or whose binary reports did not align with their reported BCage (i.e. their BCage was after the age when they first indicated a previous BC diagnosis). This resulted in an analytic sample of 3,608 individuals, 137 (3.7%) of whom reported having ever been diagnosed with BC.

We wish to assess the performance of our smoothing procedure on real data using the same three metrics as our simulation: ARISE, ARMSEw, and ARSMEo. Thus, we randomly split the sample into two halves, one of which is used to estimate the time-to-BC diagnosis curve and asses the within sample interval-censored prediction performance, and the other which is used to assess the out-of-sample prediction performance. As the number of onset evens is comparatively small, sampling variation within a single split sample could have a large influence on our results. Therefore, we performed 100 random splits of the dataset, calculating the RISE, ARMSEw, and ARMSEo each time using both the raw and smoothed (using the lnNe $BIC_s$ penalty) TB estimates. Since the true underlying population-level survival curve is unknown, we considered the K-M survival function estimated from the full dataset of exact reported BCage as the reference $S(t)$ when calculating ARISE. As in the simulation procedure, we utilized the interval and survival packages to obtain the raw TB and KM survival function estimates. This procedure provides a distribution of values for each performance metric for both raw and smoothed TB estimates. Thus, we conducted paired hypothesis tests to determine whether our smoothing method improves performance compared to the raw TB estimates.

**Results**

The results of S1 comparing the performance of smoothed versus raw TB or KM output across ARISE, ARMSEw, and ARMSEo are summarized in Figure 2 and Table 2. For the TB method, we compared the performance of three different $BIC_s$ penalties, lnN, $lnN_m$, and $lnN_e$. The results for each of the 108 parameter configurations are detailed in Supplementary Figure 3. Figure 2 shows boxplots of the simulation ARISE, ARMSEw, and ARMSEo across each of the 108 simulations, testing for a median difference less than zero using a paired Wilcoxon signed-rank test. Table 3 lists the median, minimum, and maximum values of the percent difference between the raw non-parametric methods and corresponding optimally smoothed estimation on each of the three performance metrics, along with p-values from a one-sided Wilcoxon one-sample test for a median difference less than zero. We do not assess the performance of within-sample prediction for KM estimates on RMSEw, as there are no interval-censored observations in the sample used in estimation.

On average, our proposed smoothing method performs significantly better than or equivalent to the raw TB estimation method in almost every simulation configuration considered. Notably, our method did not show significant improvement over raw TB only under a few scenarios when the study design was poor (u = 30, failing to capture most onsets) in combination with a small sample size (N < 100) and low prevalence (p = 0.1). The largest improvement of TB smoothing over raw performance appears to be when the goal is out-of-sample prediction for a highly prevalent trait and the sample size used in estimation is rather small. In most situations considered, smoothing applied to the KM estimate also shows significantly improved performance. Specifically, smoothing did not improve KM estimate performance in terms of RISE when the disease prevalence is high and the sample size is also large. Overall, the three $lnN_s$ penalty terms demonstrate similar performance. However, the $lnN_e$ penalty generally appears to show fewer individual simulation replicates where smoothing performed worse than the original TB method, thus we consider the $lnN_e$ penalty to have marginally better characteristics than either the $lnN$ or $lnN_m$ penalties.

Across most interval-censoring simulation scenarios in S1, our procedure chose a smoothing bandwidth close to that which minimized the RISE or RMSEw (Supplementary Figure S4). Sub-optimal smoothing bandwidths (chosen bandwidth smaller than optimal) were sometimes selected in scenarios when p was small (0.1) or when the study design did not match the true underlying risk period well (u = 30, u = 70). However, even with sub-optimal smoothing, we found improvement over the un-smoothed raw TB estimates.

The goal of S2 was to demonstrate the robustness of our method to handle a true underlying distribution whose density is bimodal. Assuming an underlying time-to-event distribution which is a mixture of two log-normal densities, we used simulated data to estimate raw KM and TB survival curve estimates, to which we subsequently applied our smoothing optimization algorithm. The performance results of our smoothing method in this single configuration across 500 simulation replicates are shown in Figure 3. The figure is annotated to show one-sided p-values from paired Wilcoxon tests that the mean rank of the smoothed results is less than that of corresponding raw results. We observe significantly better performance of our proposed smoothed method across all metrics except when comparing the adjusted RISE of the smoothed KM curve ($p < 2e-16$).

S3 was designed to assess the performance of bootstrapping as a method for approximating confidence intervals around our smoothed survival curve estimate. Supplementary Figure S5 shows the true underlying survival function, those estimated by raw and smoothed TB, and their 95% confidence intervals from 200 bootstrapped samples for one simulation replicate. Across 100 such simulation replicates, we calculated the proportion of times the true survival curve fell within the confidence intervals across all age bands. Results confirmed that bootstrapping performs as expected, with the average proportion of the true survival curve captured by the raw TB method being 96% (s.d. 4%) and that of the smoothed curve being 95% (s.d. 4%).

We applied our smoothing procedure to data on breast cancer from the MIDUS dataset. To assess the performance of three metrics, we performed 100 random 50-50 splits of the full dataset, each time using one half to obtain raw and smoothed TB estimates, and the other half to assess out-of-sample predictive accuracy. The distribution of these three metrics across the 100 split-sample estimates are shown in Figure 4. To conduct a formal statistical test for differences between the performance on each metric, we first conducted a Shapiro-Wilk test of normality on the difference between raw and smoothed values, which showed that only the RISE was not normally distributed ($p = 0.02$). Thus, using a paired sample Wilcoxon test for a lower mean rank of the smoothed compared to raw performance metrics, we observe a significantly improved performance ($p < 2e-16$) in terms of RISE, when the KM estimate was considered as the reference for calculating RISE. For RMSEw and RMSEo, we performed a paired sample t-test, with the alternative hypothesis that the average smoothed performance was less than that of the raw TB, again finding significantly better performance of our smoothing method over the result using raw TB (both p-values $< 2e-16$). We further used the full available sample to obtain the traditional KM, raw TB, and smoothed TB survival curve estimates, as shown in Figure 5. We can see that the raw TB, and thus the smoothed TB, results arrive at a lower endpoint for time-to-BC onset than the empirical KM curve. This can occur when relatively narrow onset event intervals happen to be observed close to the end of the curve, where there are less remining individuals contributing to the estimation who have not developed BC.

**Discussion**

We have proposed a smoothing algorithm for reducing the overfitting problem encountered in traditional NPMLE (TB and KM) estimation and provide an R package for convenient implementation (https://github.com/tubbsjd/SISE). Our algorithm appears to impose little additional computational burden over existing implementations of survival curve estimation, with a runtime increase of only about one second when performing smoothed TB estimation on the complete MIDUS breast cancer dataset.

Through extensive simulations across various combinations of assumed disease onset characteristics and study designs, we have shown that our proposed smoothing method applied to traditional raw estimates from TB estimation significantly improves performance on a number of important metrics, namely the divergence from the true population-level curve, as well as both within- and out-of-sample prediction error. In applying the same procedure to the

KM-estimated procedure using exact and right-censored observations, we find smoothing improves performance across all metrics except in those simulated scenarios when the disease has a high lifetime prevalence and the sample size is relatively large. This is to be expected, since the number and the quality of observations are already sufficient for raw TB or KM estimation to provide an accurate estimate.

Although we carried out extensive simulations across a large number of common disease processes and study design scenarios, there may still be some unknown cases where our method does not demonstrate improved performance over the standard TB method for censored data. However, we show that our method is robust to simulations where the true time-to-event distribution is bimodal and that confidence interval approximations can be obtained through a bootstrapping procedure. Furthermore, we show that in practice, our smoothing algorithm is also able to improve estimation accuracy and predictive performance in a real dataset of interval-censored breast cancer onset ages from the MIDUS longitudinal study.

Our proposed method applies a normal kernel regression smoother to the estimated density with a window size chosen to minimize a BIC-type loss function of the resulting estimate. We opt for NW normal kernel regression mainly for the sake of simplicity and computational convenience, so further improvement could potentially be made by choosing some locally adaptive smoothing methods such as those based on wavelets and spines.

Our $BIC_s$ penalizes the data likelihood by a measure of the degree of complexity, which is the number of "turning points" (strict local extrema) on the estimated density curve, weighted by the logarithm of a sample size-informed quantity. The strength of smoothing decreases when data informativeness (as measured by $N_e$) increases. The number of "turning points" is not the only possibly suitable measure. When trying to estimate the effective number of parameters, other metrics to quantify variability of a function may be considered, such as auto-correlation Lipschitz Constant and the number inflexion points. We acknowledge the prospects of generalizing variability-penalizing smoothing to broader areas of training, optimization and model selection in the field of statistical machine learning.

In theory, our variability-based modification can be transferred to other IC, such as AIC. Others have acknowledged that AIC typically overfits by introducing a sub-optimal penalty when the sample size is small, whereas BIC has a substantial risk of selecting a very bad model when sample size is much larger than the square of the number of parameters (Burnham & Anderson, 2016; Hurvich et al., 1998). As our contexts focus more on overfitting due to small sample size, we prefer BIC-type over AIC-type in our procedure. Nevertheless, we do not rule out the possibility of modifying other types of IC for a different setting or scenario.

Strictly parametric estimation can lead to biased estimation when incorrect assumptions are made about the shape of the underlying distribution. Thus, many researchers have adopted the first approach discussed earlier, i.e. moving to more valid semi-parametric approaches by allowing the assumptions to be more flexible, especially as available data increases. The increasingly popular diffusion estimator (Botev et al., 2010) is one such example. The increase in available computational resources has also promoted the popularity of similar Gaussian Mixture Models. Future work may compare the performance of our current procedure to one that adopts a Finite Mixtures framework towards survival & density function estimation from censored data.

Overall, our method contributes to the field of non-parametric survival analysis by achieving improved accuracy and predictive ability, especially when available survival data is limited by censoring, small sample size, low prevalence, or few repeated measurements. We have demonstrated that smoothing the NPMLE by complexity-based penalized likelihood is a simple but powerful approach for improving density estimation, and we believe that this approach is likely applicable or generalizable to any raw result of the EM algorithm.

# Tables

**Table 1. An illustrative hypothetical dataset of heart attack history**

| | *Disease status and age (Z, T)* | | | *Compressed data (L, R)* |
|---|---|---|---|---|
| *Individual* | 1st obs. | 2nd obs. | 3rd obs. | |
| 1 | 0, 38 | 1, 60 | 1, 70 | (38, 60) |
| 2 | 0, 41 | 1, 48 | 1, 55 | (41, 48) |
| 3 | 0, 35 | 0, 44 | 0, 62 | (62, +∞) |
| 4 | 1, 36 | 1, 42 | 1, 48 | (0, 36) |
| ... | … | | | … |

1 = previous heart attack reported; otherwise, 0

**Table 2. Summary of Simulation Configuration Parameters**

| | | PARAMETER VALUES | | | | | |
|---|---|---|---|---|---|---|---|
| SIMULATION | DESCRIPTION | N | u | p | m | $N_s$ | $N_{rep}$ |
| S1 | Compare all 24 possible configurations of parameters (N, u, p, m). Compare performance of three $N_s$ alternatives | 50<br>100<br>1,000<br>5,000 | 30<br>50<br>70 | 0.1<br>0.5<br>1.0 | 2<br>4<br>6 | $\ln N_e$<br>$\ln N$<br>$\ln N_m$ | 100 (each scenario) |
| S2 | Test sensitivity of smoothing method for bimodal time-to-event distributions. | 500 | Mixture<br>u = 30,<br>u = 60 | 0.75 | 2 | $\ln N_e$ | 500 |
| S3 | Tests the performance of bootstrapping for estimating smoothed confidence intervals. | 100 | 50 | 1 | 6 | $\ln N_e$ | 100 |

**Table 3. Median (Minimum, Maximum) Values for the Percent Change (Compared to Raw Estimate) for Performance Indices Across 108 Configurations for Smoothed TB and KM using different BIC penalty terms**

| Method | $N_s$ | $\Delta$ARISE | p-value | $\Delta$ARMSEw | p-value | $\Delta$ARMSEo | p-value |
|---|---|---|---|---|---|---|---|
| TB | $\ln N$ | -0.22(-0.49,0.04) | 1e-19 | -0.15(-0.40,0.00) | 9e-20 | -0.12(-0.22,0.00) | 9e-20 |
| TB | $\ln N_m$ | -0.22(-0.49,0.04) | 1e-19 | -0.16(-0.41,0.00) | 9e-20 | -0.12(-0.23,0.00) | 9e-20 |
| TB | $\ln N_e$ | -0.21(-0.49,0.04) | 1e-19 | -0.15(-0.38,0.00) | 9e-20 | -0.11(-0.22,0.00) | 9e-20 |
| KM | $\ln N_e$ | 0.07(-0.15,0.13) | 2e-13 | | | -0.02(-0.17,0.00) | 4e-19 |

P-values are from a Wilcoxon signed-rank test with the alternative hypothesis that the smoothed result is lower than that of the corresponding raw result. TB = Turnbull Estimation, KM = Kaplan-Meier Estimation.

# Figures

**Figure 1. The behavior of BIC$_s$ components under simulation**

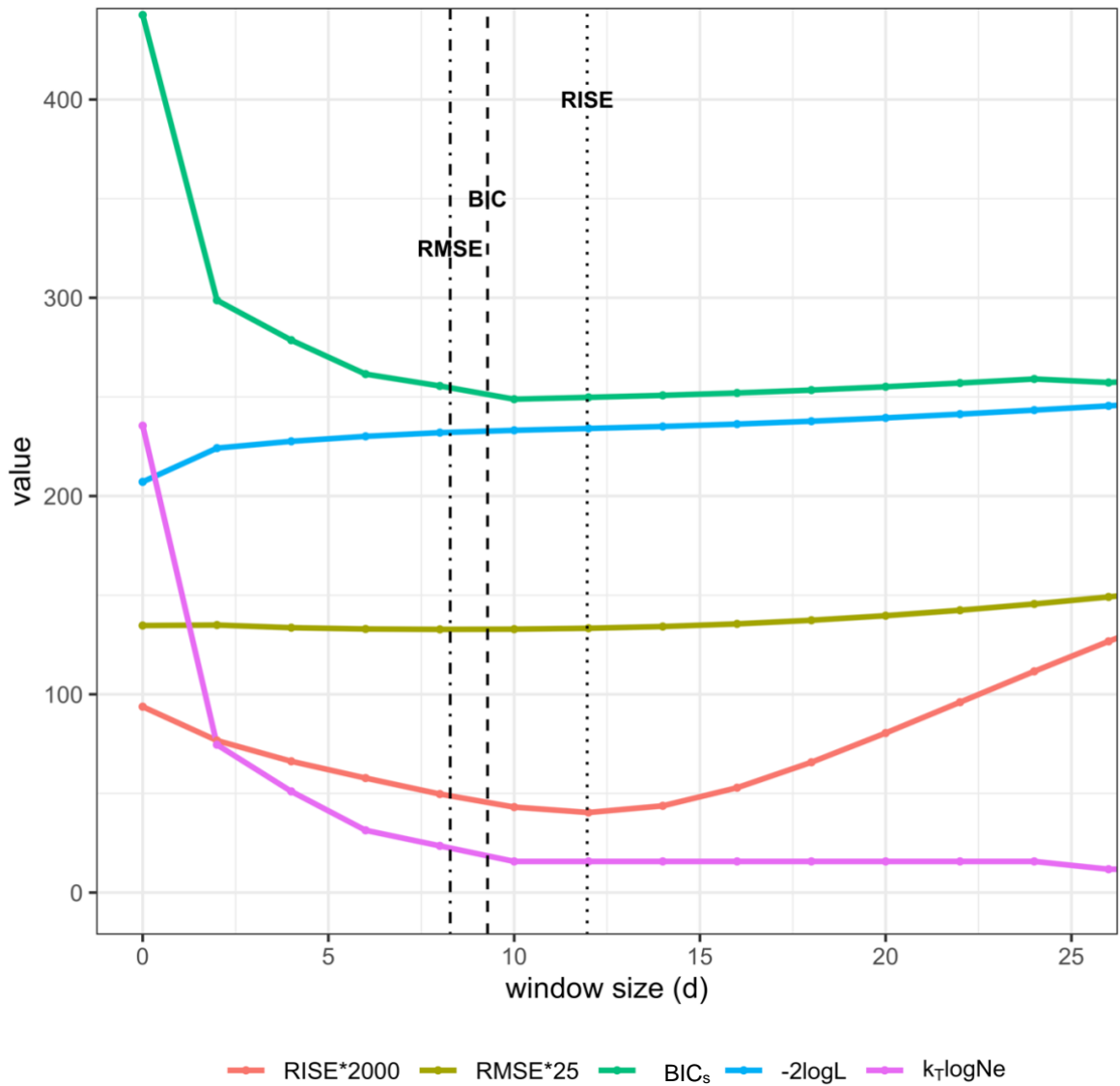

The behavior of BIC$_s$ components and performance metrics with increasing window size in a single simulation with N = 100, u = 50, p = 1, and m = 6. RISE and RMSE are scaled so that their behaviour can be clearly shown in the plot. Note that a window size of 0 corresponds to the original Turnbull estimate.

**Figure 2. Simulation Results Averaged Within 24 Configurations**

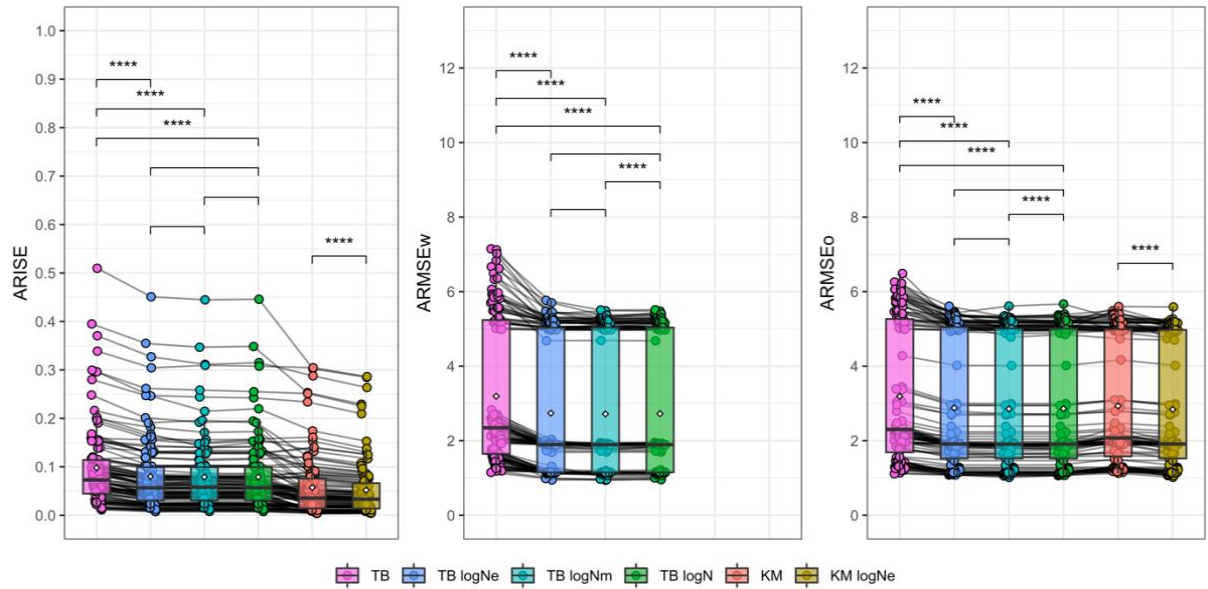

Boxplots showing values of ARISE, ARMSEw, and ARMSEo performance metrics across the 108 simulation configurations, each with 100 replicates, for raw and smoothed estimates. Points represent the values for individual configurations, with lines connecting values from the same configuration across methods. Stars indicate the p-value of a Wilcoxon signed-rank test where the alternative hypothesis is that the median value of the smoothed curve is less than that of the raw curve. When comparing lnNs, lnN was considered as the reference, except when comparing lnNe and lnNm, where the latter was considered the reference. (* $p \leq 0.05$; *** $p \leq 0.001$; **** $p \leq 0.0001$)

**Figure 3. Performance of Smoothing when True Distribution Density is Bimodal**

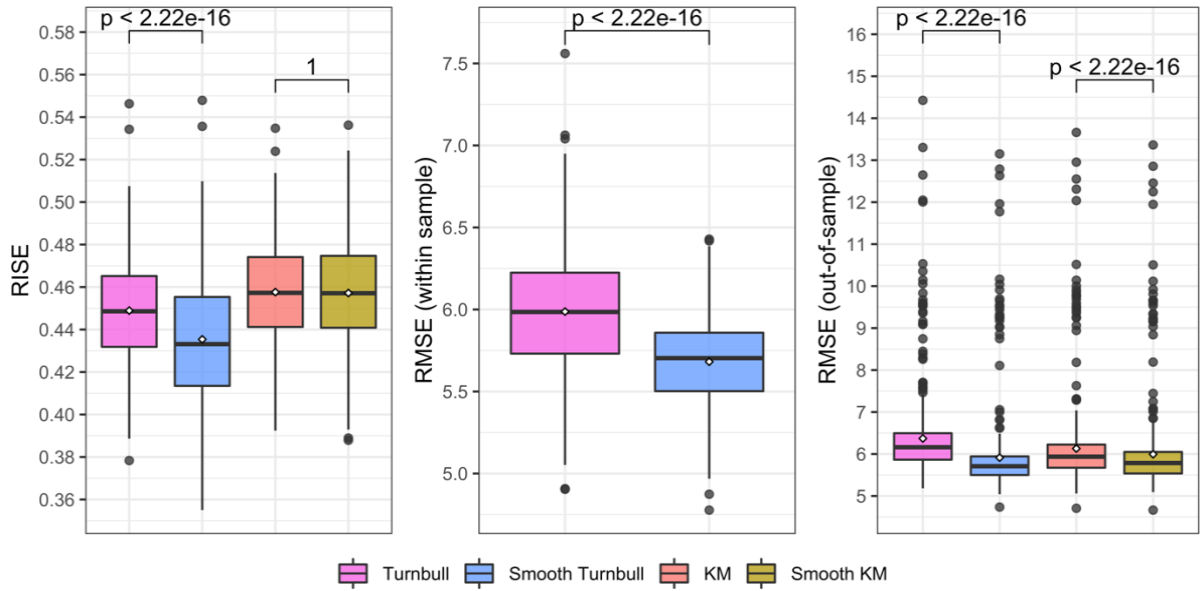

Box-and-whisker plots indicate the distribution of simulation results across three performance metrics, annotated to show p-values from paired Wilcoxon tests where the alternative hypothesis is that the smoothed result is smaller than that of the raw result, which show significantly better performance of our proposed smoothed method across all metrics except when comparing the adjusted RISE of the smoothed KM curve.

**Figure 4. Distribution of 3 Performance Metric Estimates Across 100 Random Split-Validation Replicates in Breast Cancer Data**

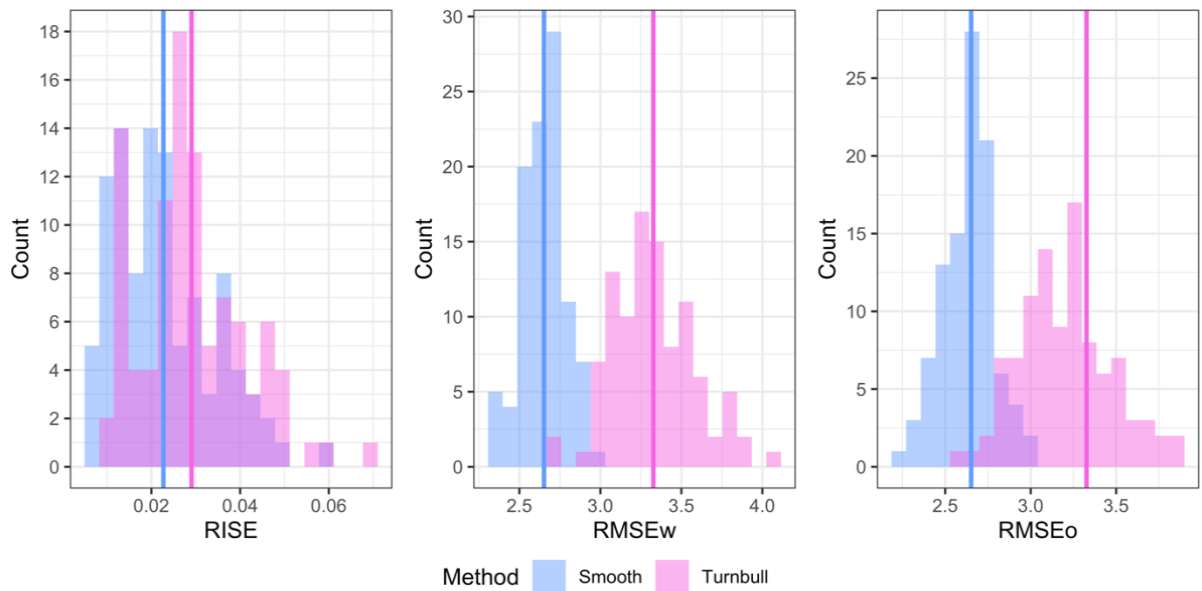

Distribution of performance metrics from real data on breast cancer diagnosis from 100 random 50-50 splits of the full dataset. Each time, one half was used to obtain raw and smoothed Turnbull estimates, which were compared to the KM estimate from the full sample by the adjusted RMISE. RMSEw was used to compare the true simulated onset age and that imputed from the estimated onset probability density curve within the same sample of interval censored observations. Similarly, the second half sample was used to assess the out-of-sample prediction accuracy using RMSEo. The means for each metric within each method are shown as vertical lines on the histograms.

**Figure 5. Time to Breast Cancer Diagnosis Estimated in the MIDUS Dataset**

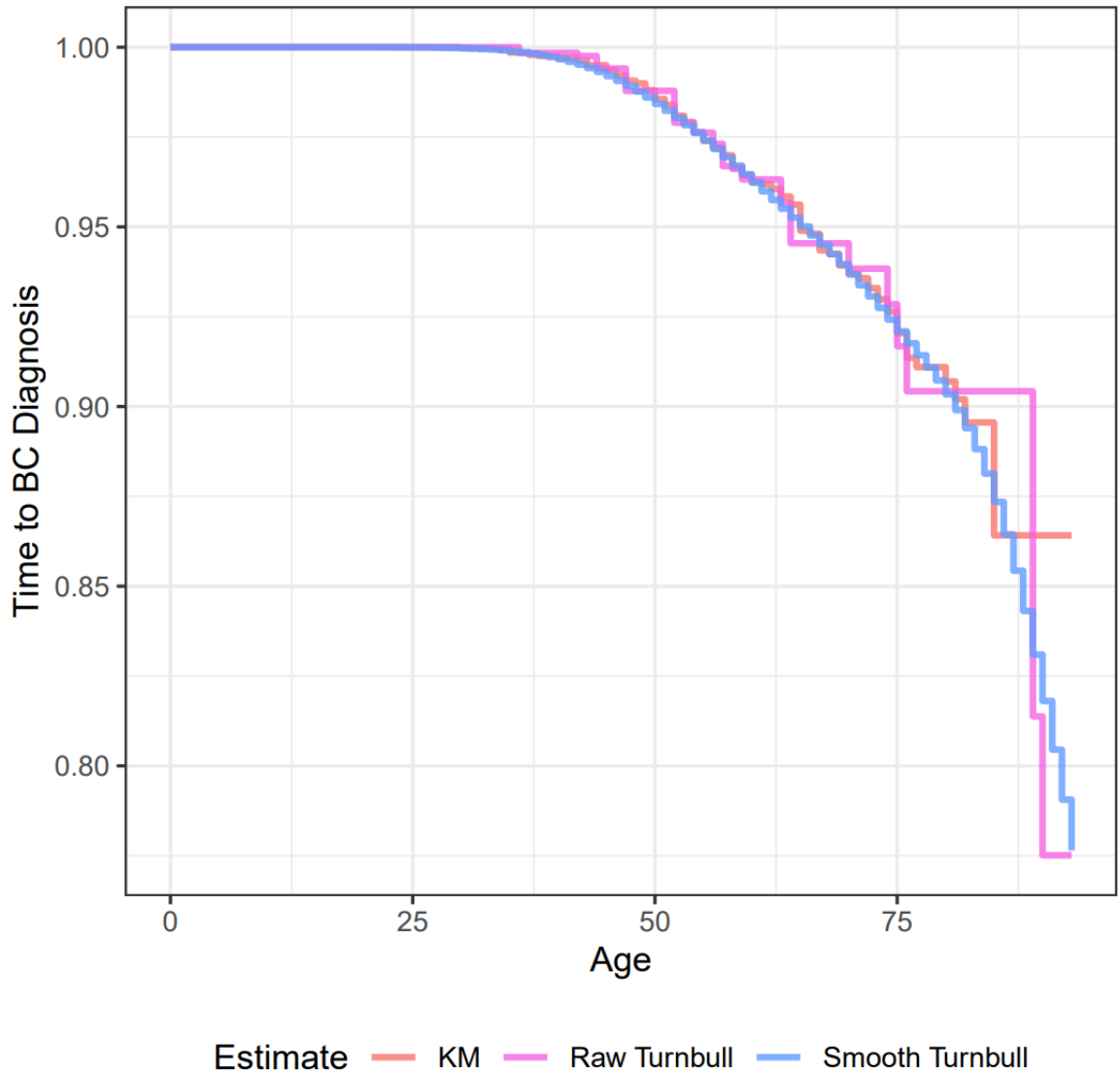

Time to breast cancer diagnosis estimated in the MIDUS dataset using traditional KM estimation with exact and right censored observations, raw Turnbull using interval censored observations, and the result from applying our smoothing algorithm to the Turnbull estimate.

# Supplementary Material

**Properties of $N_e$**

Here, we outline some intuitive properties of $N_e$:

1) $N_e \leq N$. In particular, when all data are exact observations, all $n_i = 1$, $N_e = N$. Larger $n_i$ are associated with smaller $|R_i - L_i|$, i.e. lower degree of censoring.

2) Particularly, when $R_i = B^L$, $n_i = 1$; $R_i = B^R$, $n_i = 0$. And when $L_i = B^R$, $n_i = p$; $L_i = B^L$, $n_i = 0$.

3) Larger $N_e$ is associated with larger m, i.e. denser observations. Larger $N_e$ is associated with larger lifetime prevalence p.

4) When our sampling scheme i.e. distribution of $T_{ij}$ matches (overlaps) well with the true truncated density $f_X(x; B)$, e.g. $ET_{ij}^k \approx EX^k, k = 1,2 \ldots$, $\min_{ij} T_{ij} \approx B_L$, $\max_{ij} T_{ij} \approx B_R$, then $N_e$ is larger.

These properties suggest that $N_e$ may be a rough indicator of the quality of the sampling scheme which generated a given dataset. Below we plot the $N_e$ versus m, p, and u to support properties 3 and 4 above.

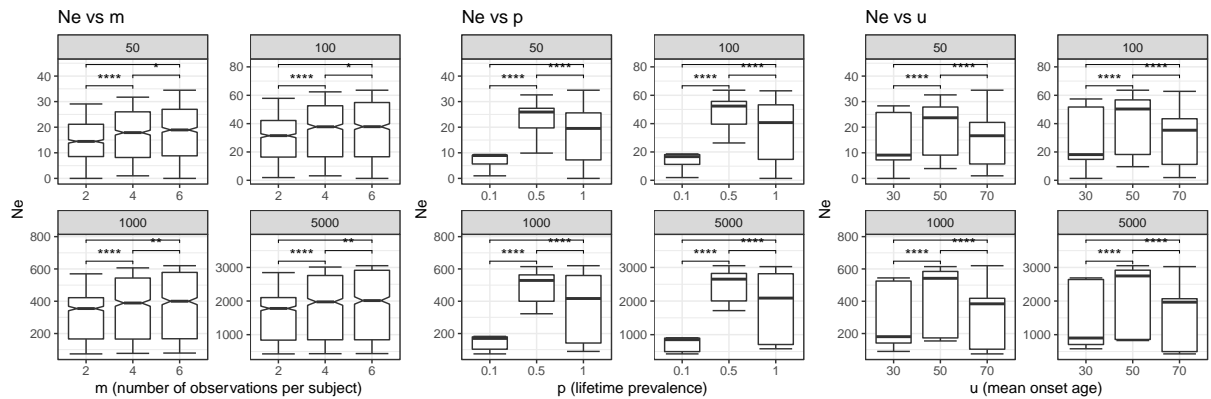

# Figure S1. Drawbacks of NP-MLE (Turnbull's method estimation)

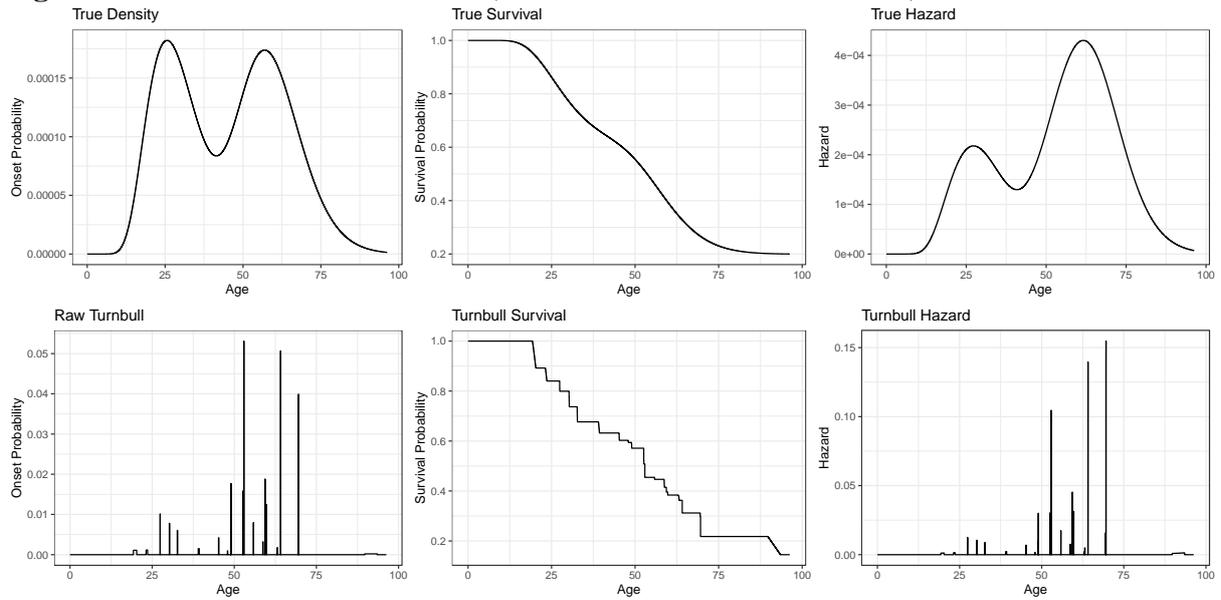

Given a true underlying distribution constituting a mixture of two log-normals, we compare the true density, survival and hazard functions to those estimated by Turnbull's method using simulated interval-censored data.

**Figure S2. Comparison of Under-smoothing, Optimized smoothing, and Over-smoothing**

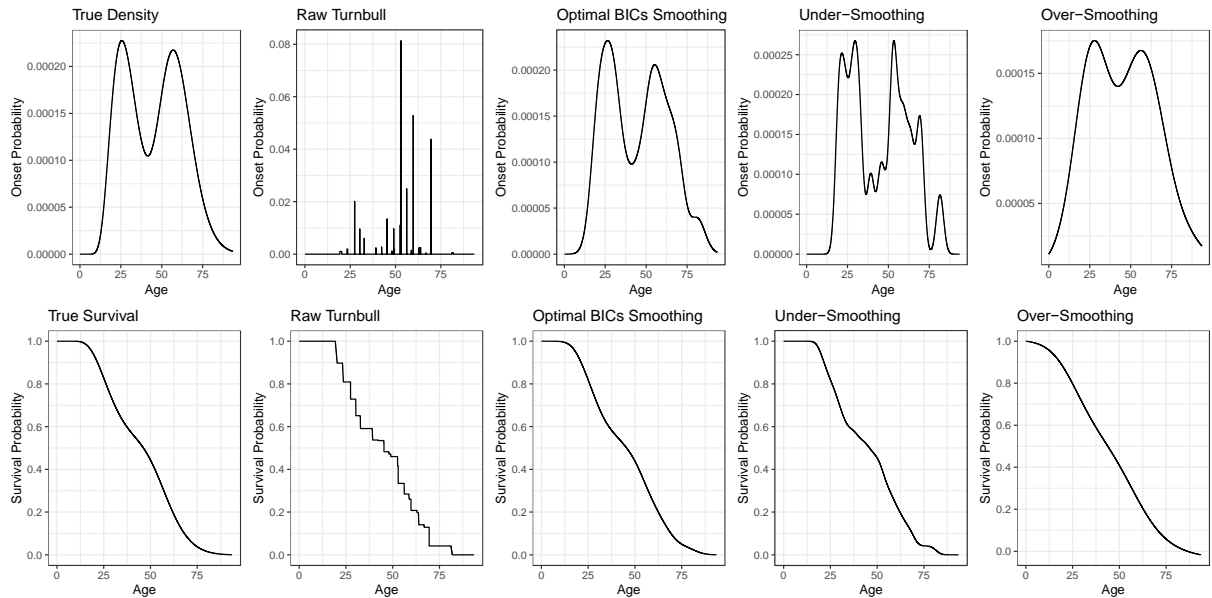

Given a true underlying distribution constituting a mixture of two log-normals, we compare the true density and survival functions to those estimated by Turnbull's method using simulated interval-censored data. We also compare these to the results after selecting an optimal bandwidth using the logNe-based BICs penalty in our proposed algorithm. We also show the results of under-smoothing (using half the chosen bandwidth) and over-smoothing (using double the chosen bandwidth).

# Figure S3. Detailed Results for Each of 108 Simulated Scenarios.

These figures show the detailed results for each of the 108 simulated scenarios across the three performance metrics (RISE, RMSEw, and RMSEo).

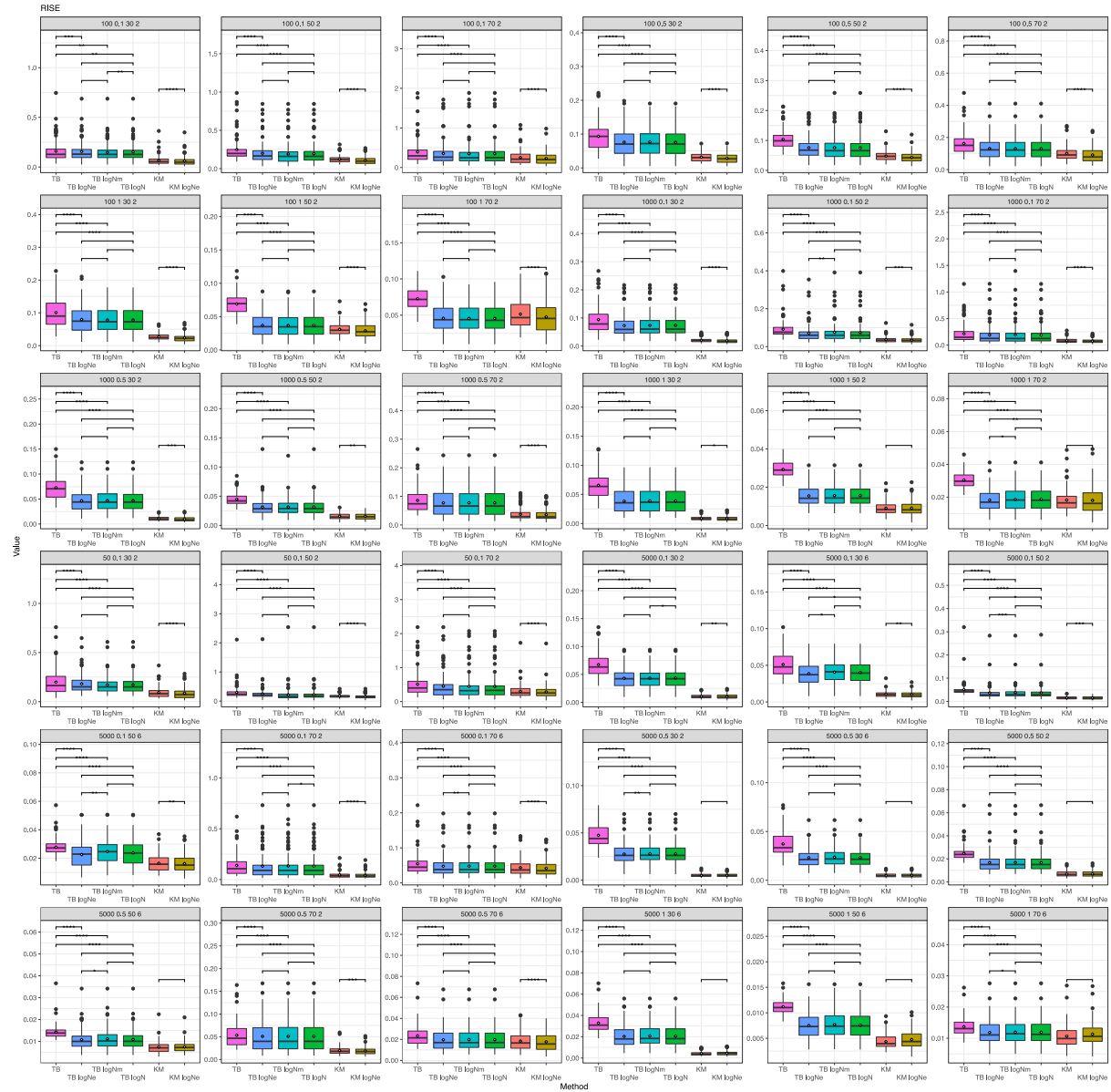

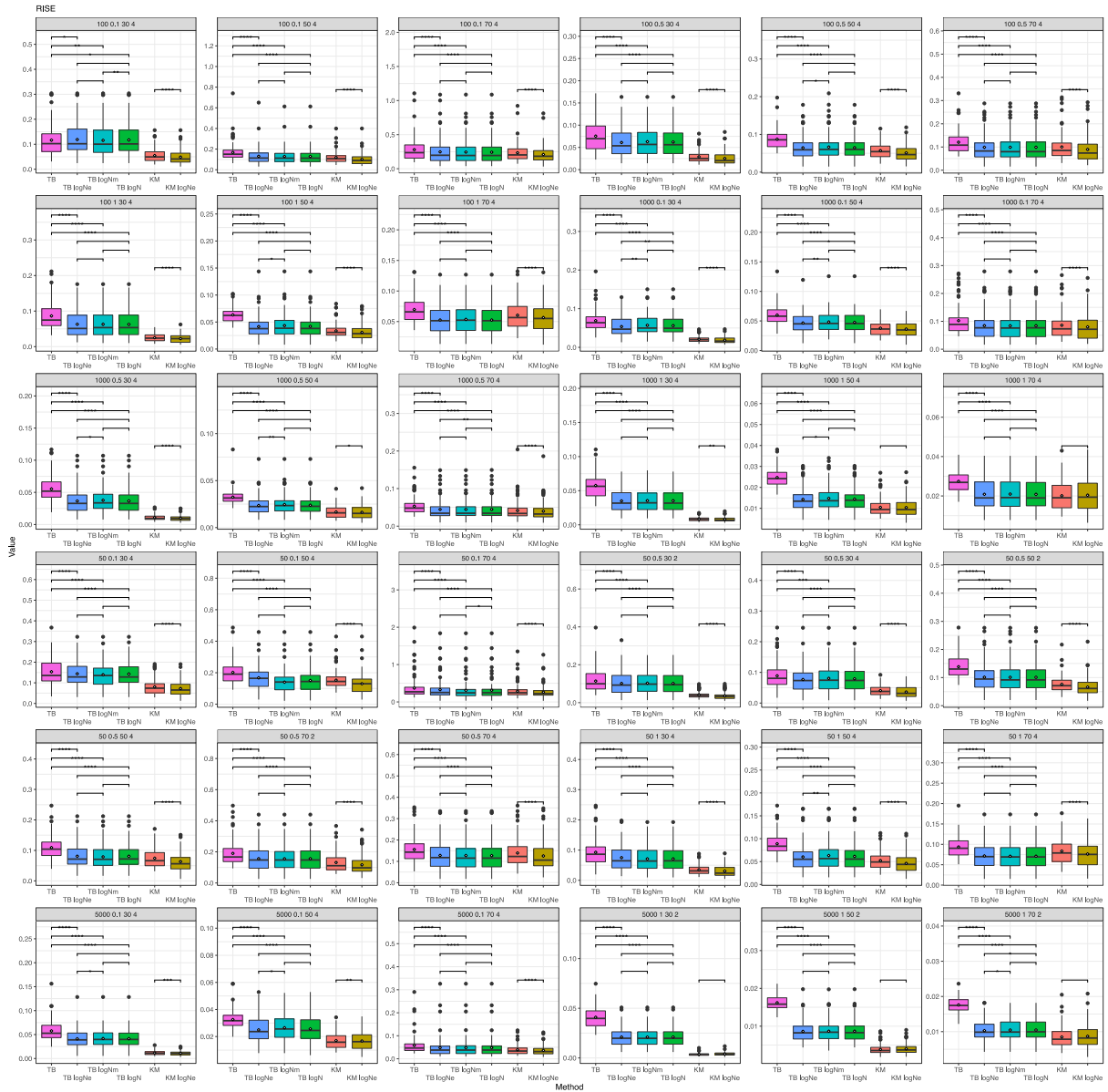

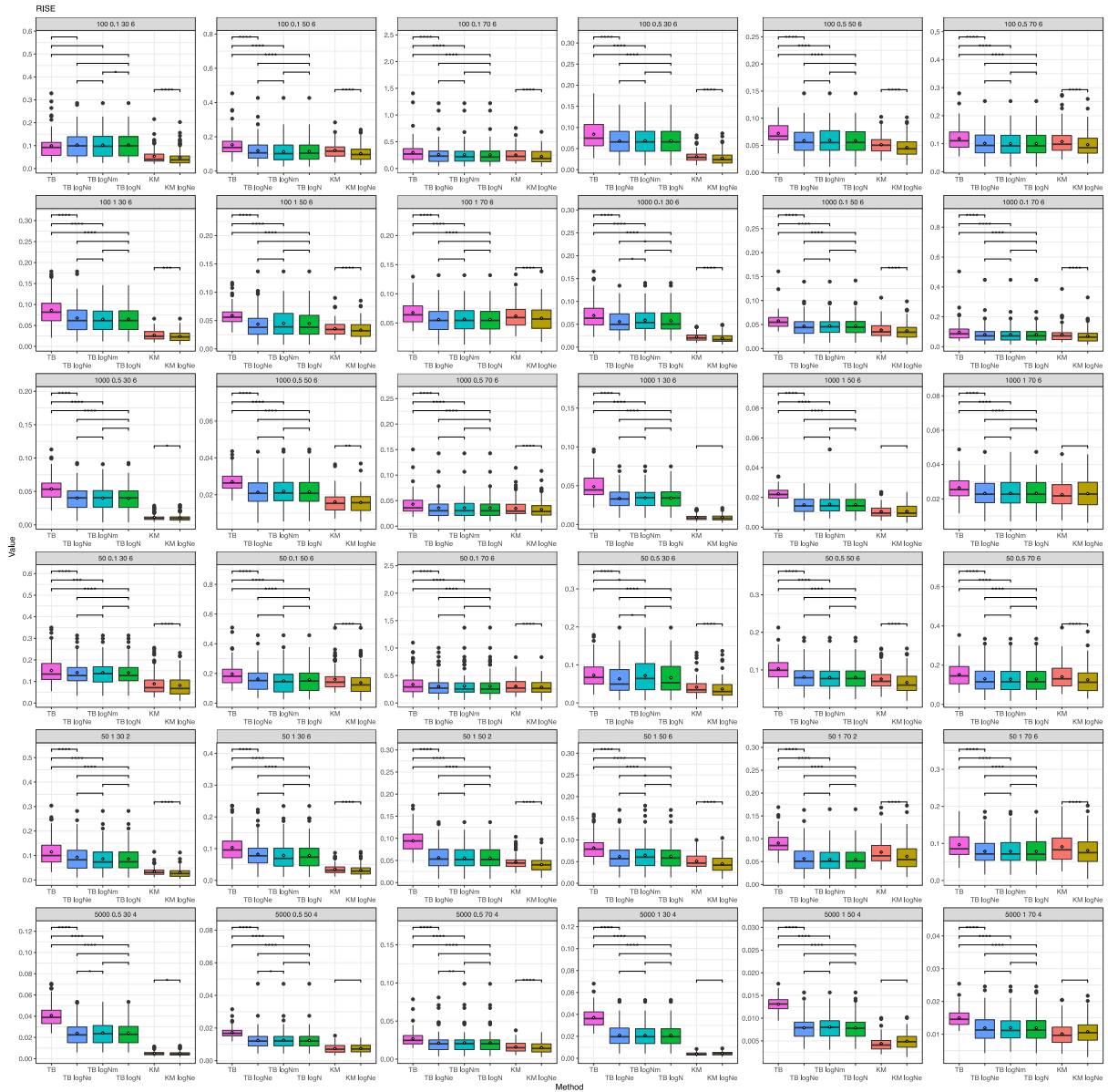

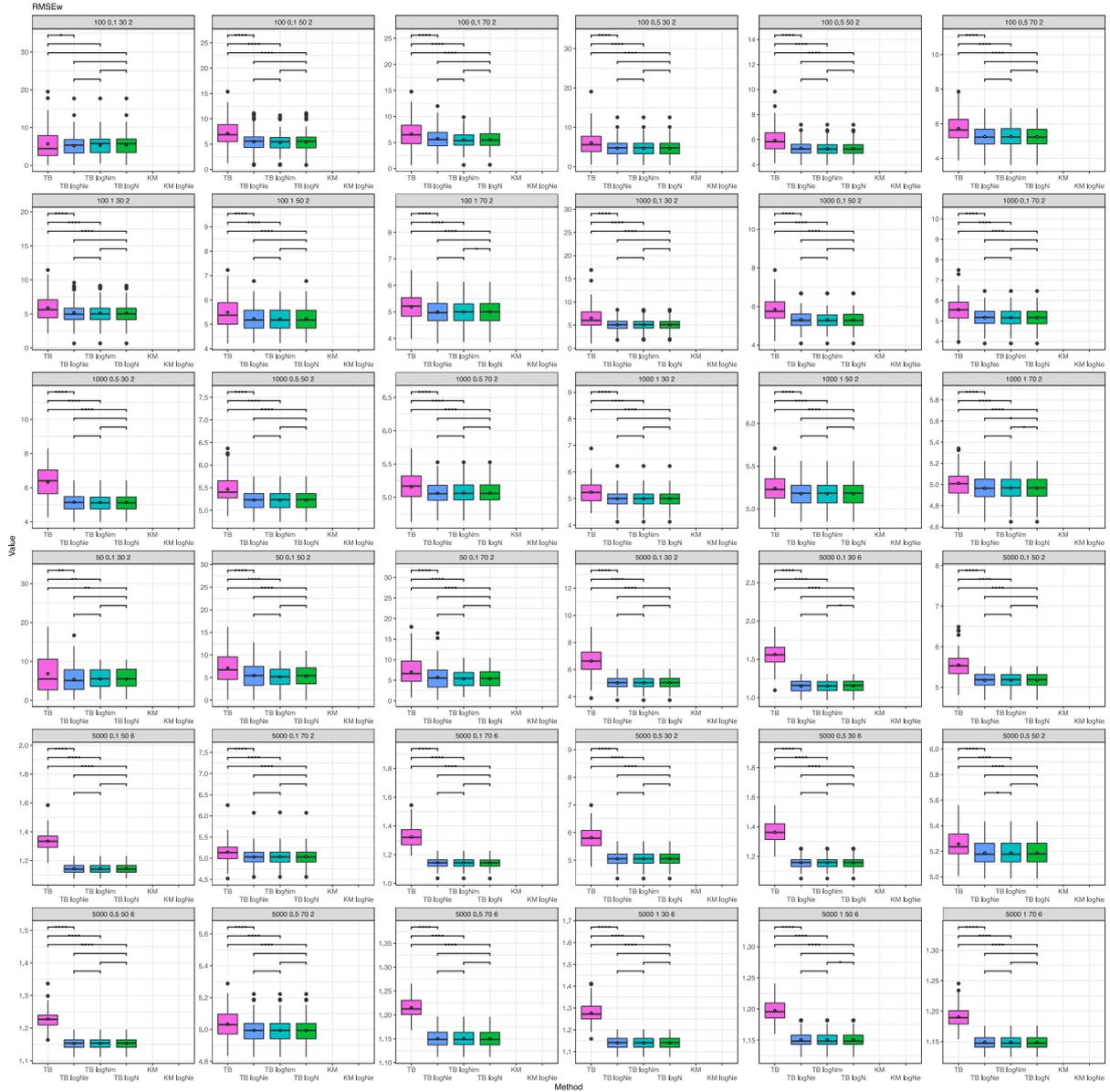

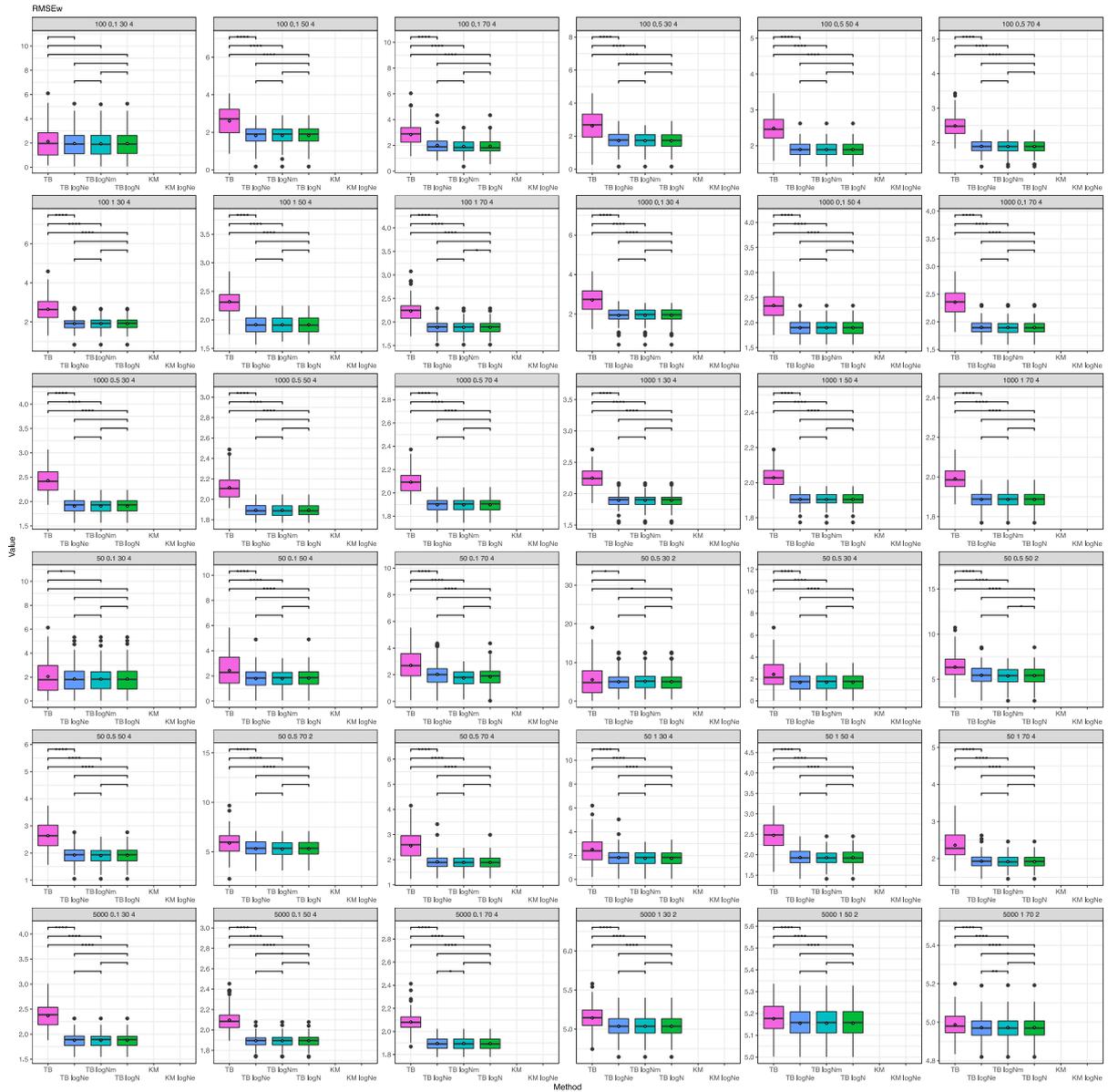

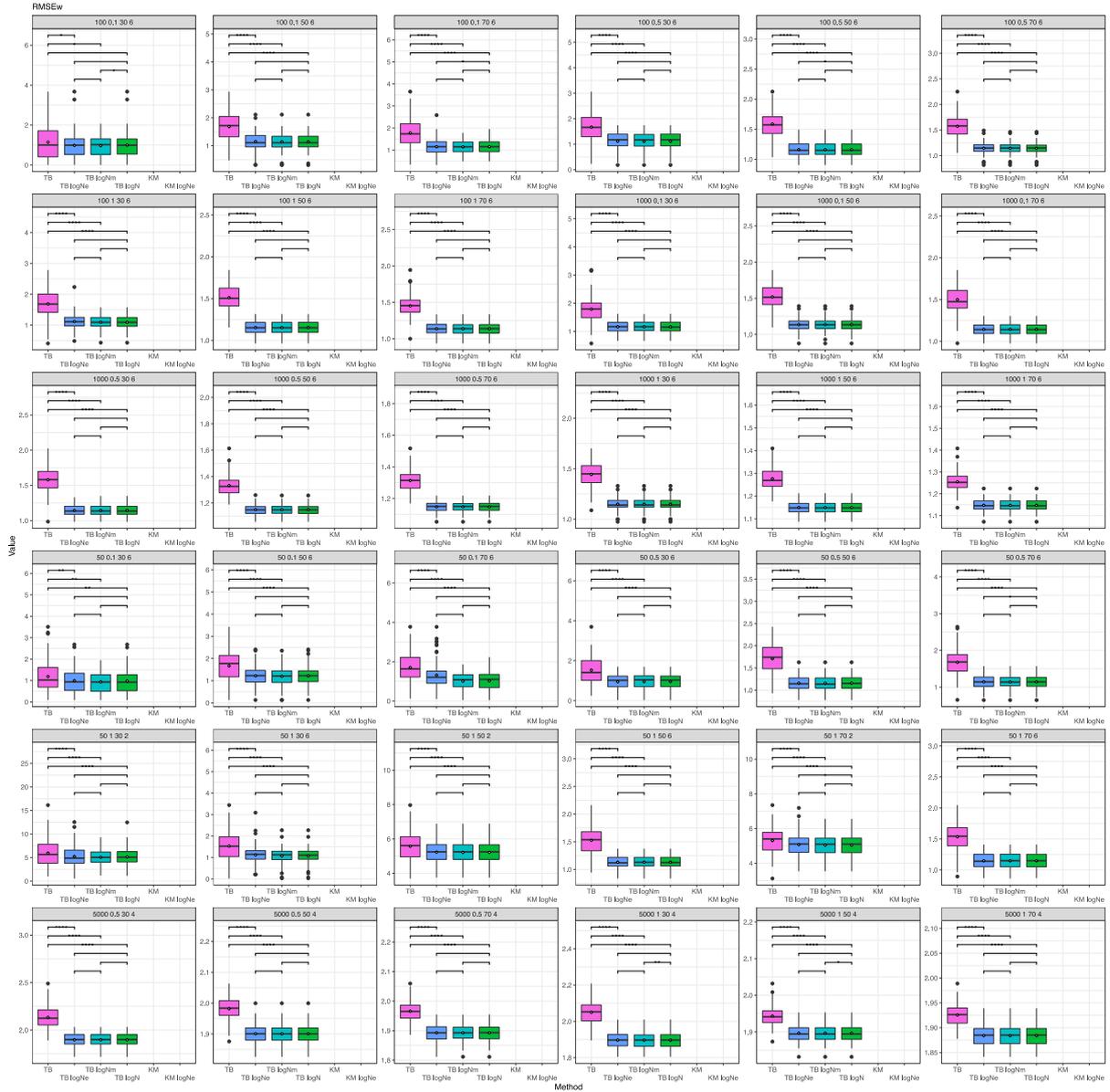

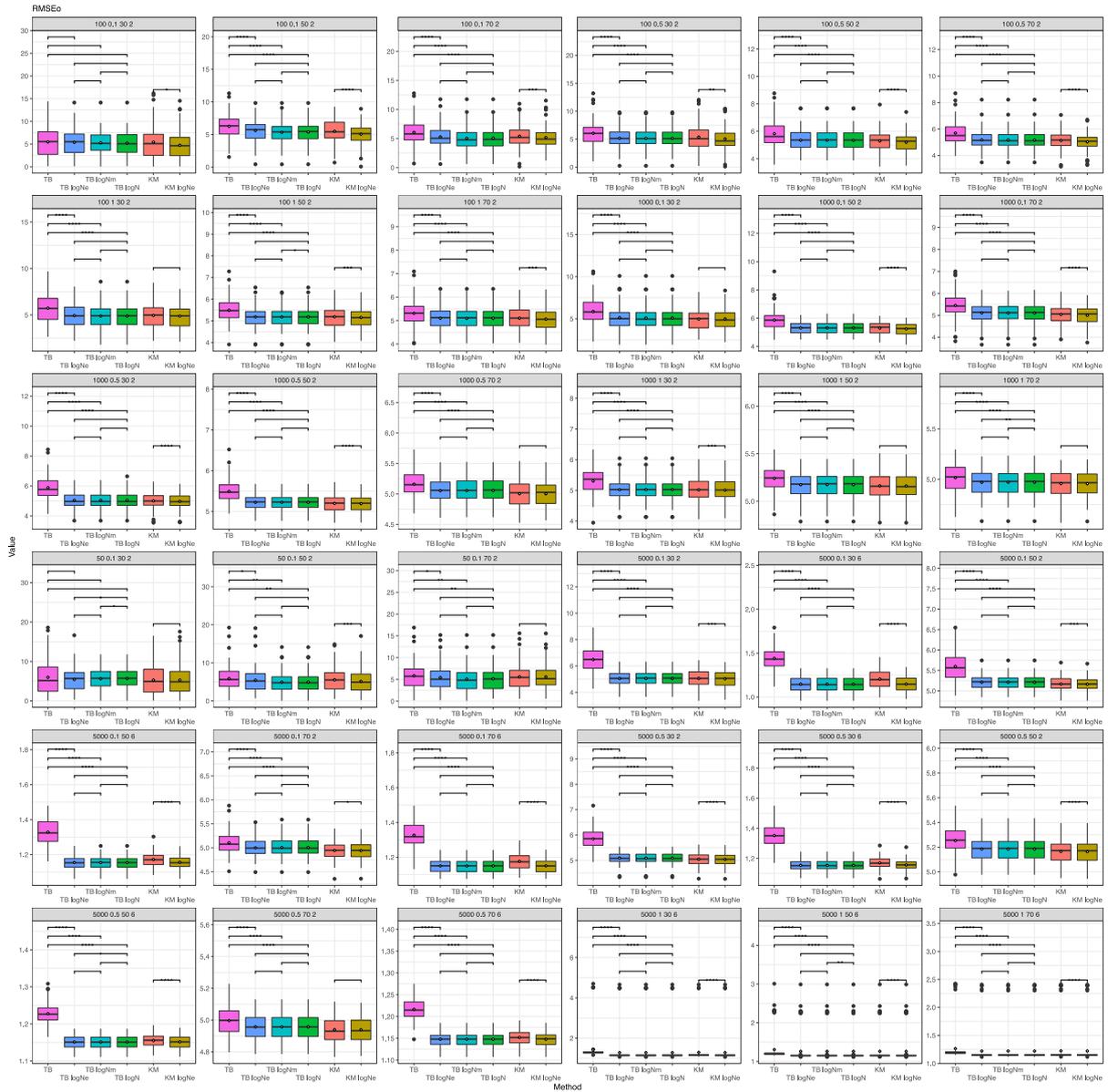

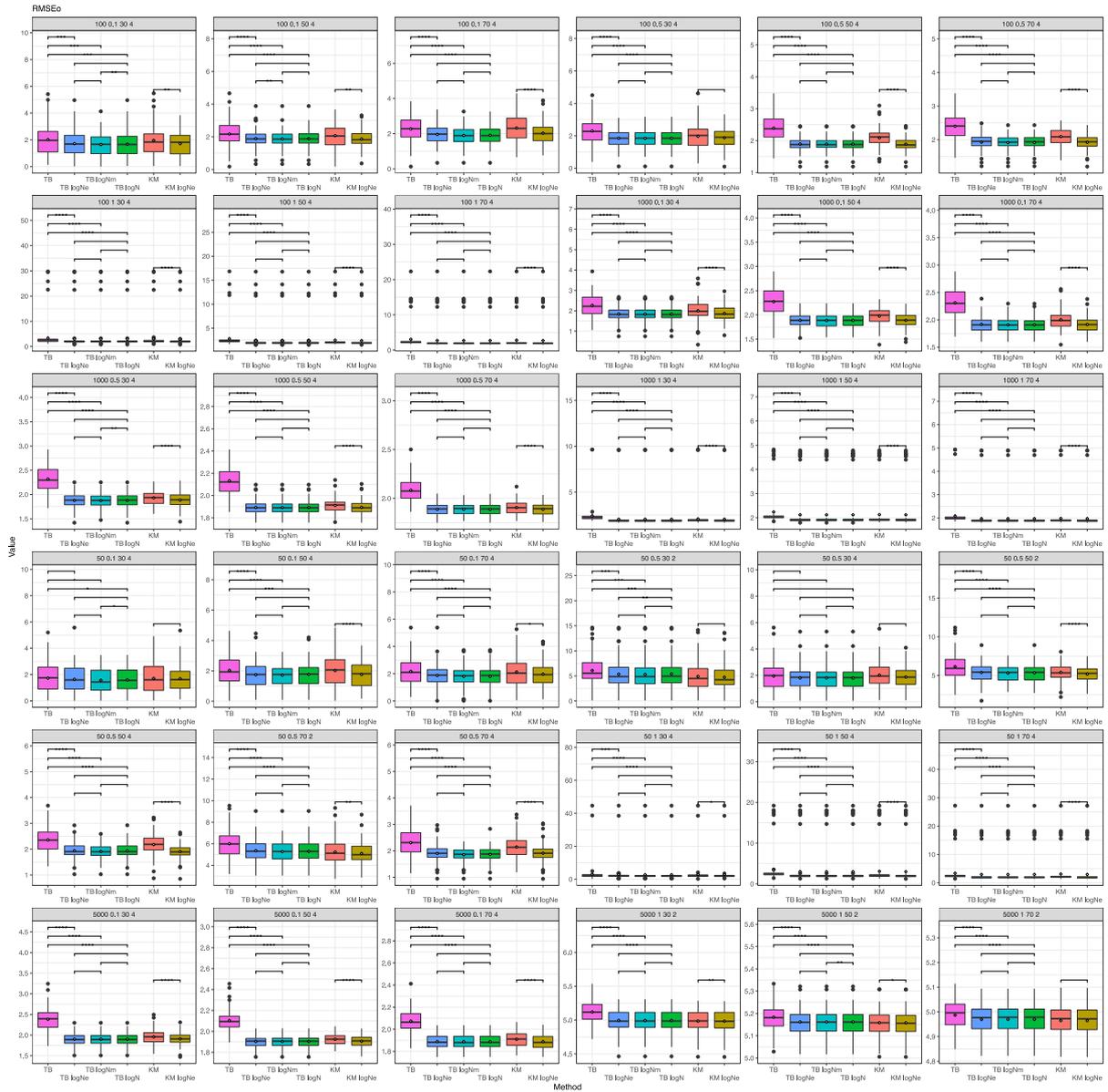

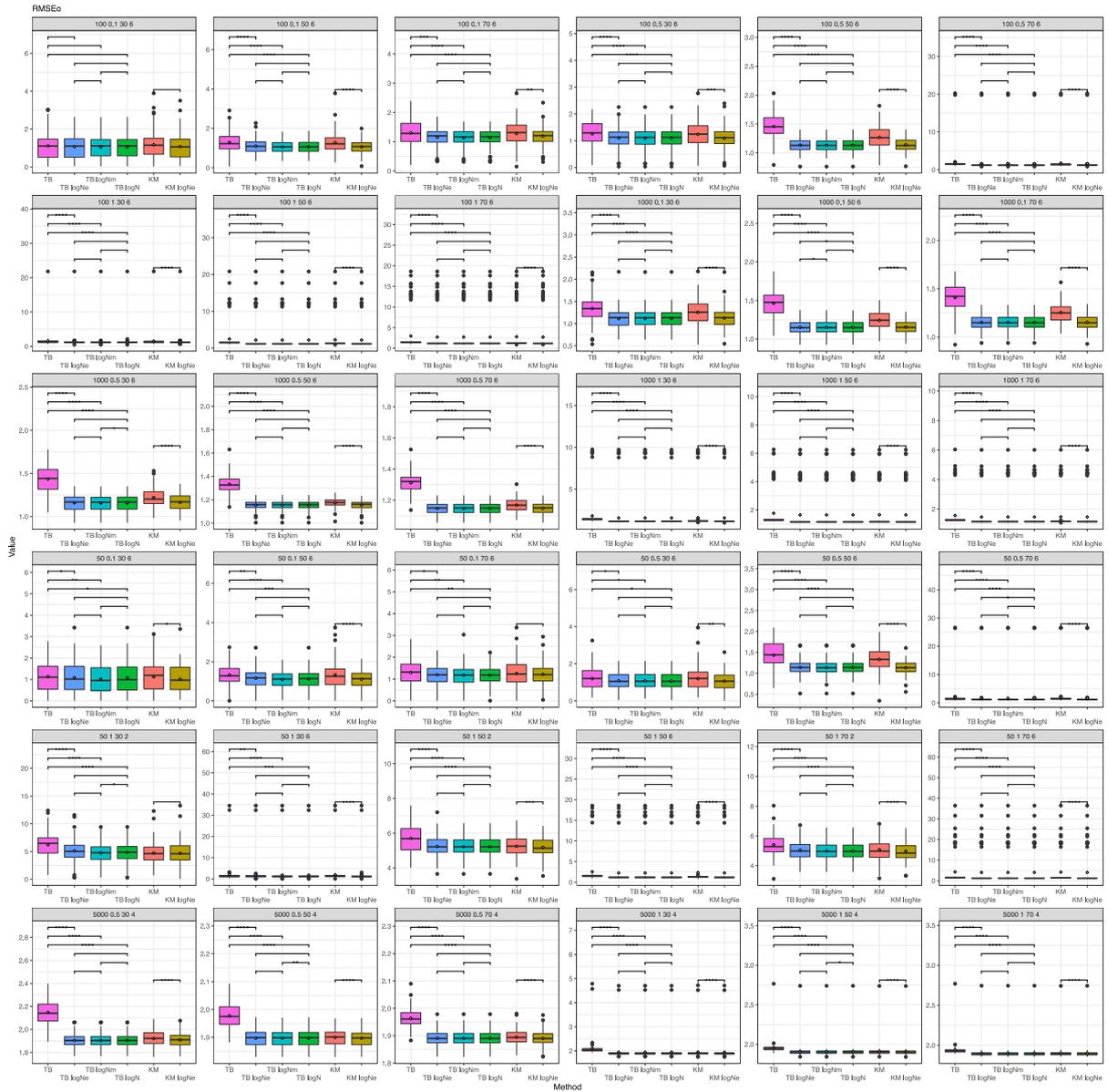

**Figure S4. The chosen vs optimal (lowest RISE or RMSEw) smoothing bandwidth**

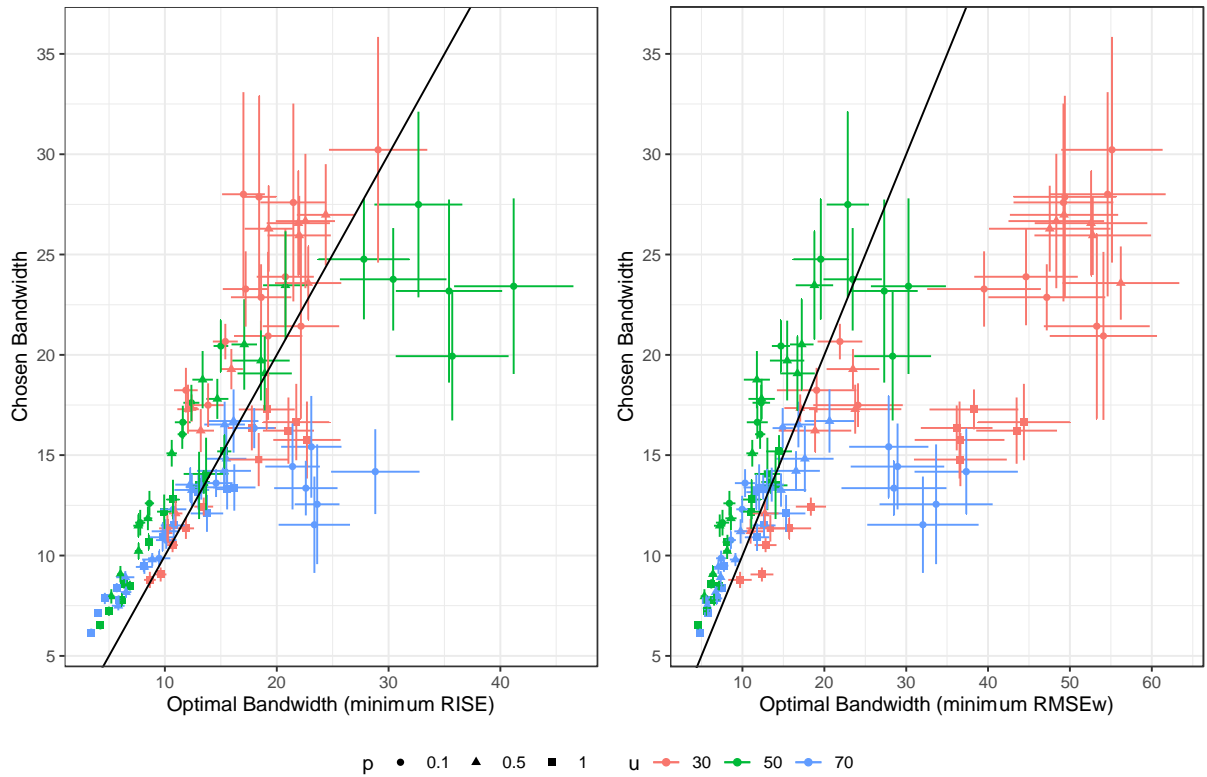

Here we show the bandwidth chosen by our method using the logNe-based $BIC_s$ penalty on the raw TB estimates versus the optimal bandwidth which minimizes either the RISE (left panel) or the RMSEw (right panel) across 108 configurations with 100 simulation replicates each. Points represent the mean bandwidth for a given configuration, with error bars showing the standard errors of the estimates. The black line represents exact concordance between the chosen and optimal bandwidths. We see that under most simulated scenarios, the chosen bandwidth is close to the optimal. However, as expected, when the sampling scheme is not optimized to the true mean onset (u = 30 or u = 70), the chosen bandwidth is under-estimated to optimize within-sample prediction.

**Figure S5. Bootstrapped Confidence Intervals from a Single Simulation Replicate**

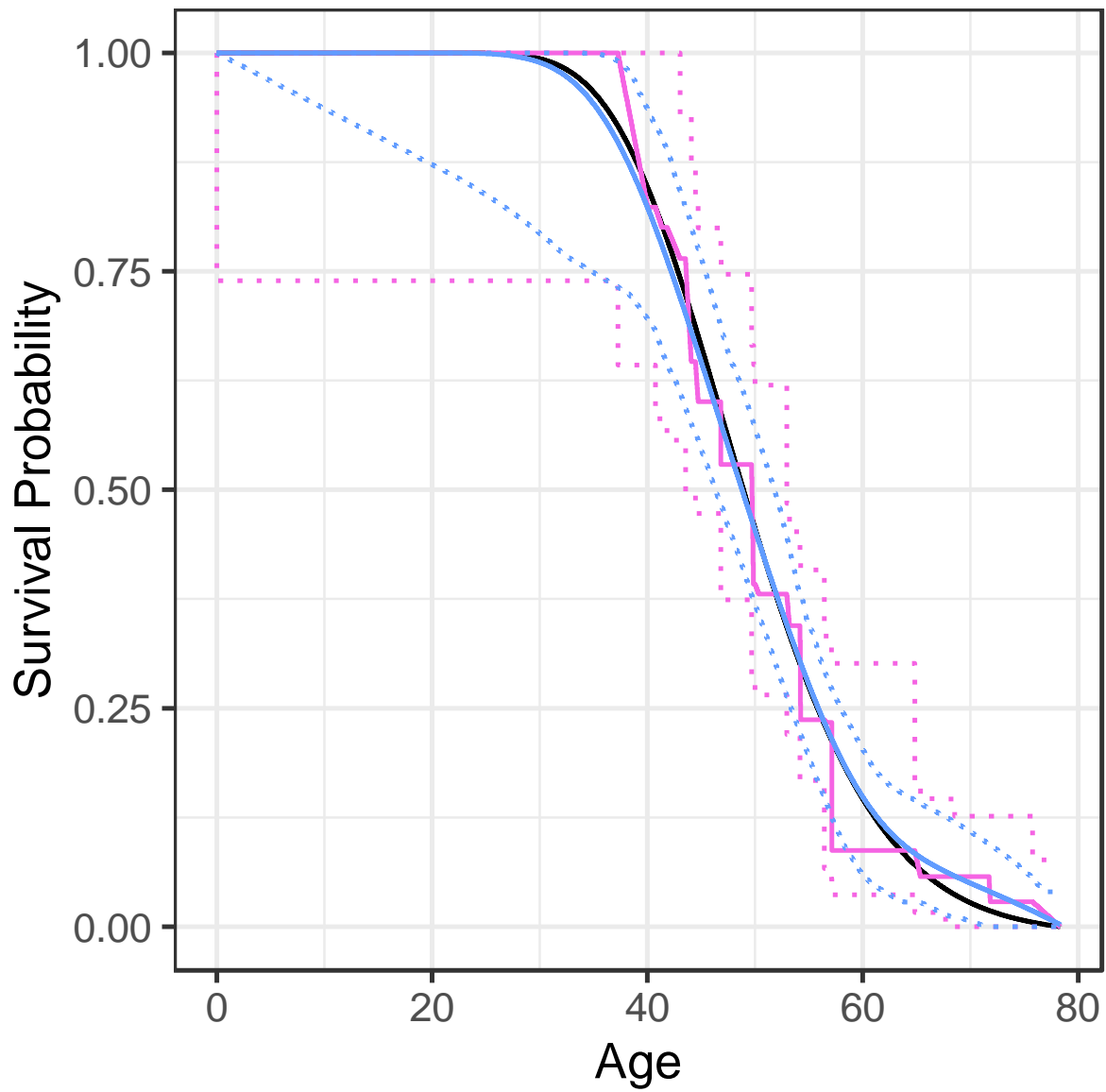

The solid black line represents the assumed true survival function, whereas the solid purple and blue lines represent the raw and smoothed Turnbull survival estimator, respectively. The corresponding dotted lines represent the upper and lower 95% confidence intervals approximated from 200 bootstrapped samples.